\documentclass[11pt]{article}
\usepackage{comment}
\usepackage[margin=2.5cm]{geometry}
\usepackage[toc,page]{appendix}
\usepackage{amsmath,amsfonts,breqn,comment,cancel,mathtools,amssymb,extarrows,mathtools,graphicx,subfigure,setspace,bbold,physics}
\usepackage{xr-hyper} 
\usepackage{cite}
\usepackage{slashed}
\usepackage[dvipsnames]{xcolor}
\usepackage{hyperref}
\hypersetup{colorlinks=true, linkcolor=blue, citecolor=magenta, linktoc=page}
\usepackage{url}
\usepackage{tikz}
\usepackage{color}
\usepackage{arydshln,leftidx,mathtools}
\usetikzlibrary{decorations.pathreplacing}
\numberwithin{equation}{section}
\newcommand{\be}{\begin{equation}}
	\newcommand{\bea}{\begin{eqnarray}}
		\newcommand{\eea}{\end{eqnarray}}
	\newcommand{\ba}{\begin{align}}
		\newcommand{\ea}{\end{align}}
	\newcommand{\ee}{\end{equation}}

\usepackage{chngcntr}
\begin{document}
	\onehalfspacing
\begin{titlepage}
	\thispagestyle{empty}

%
	
	\vspace{.4cm}
	\begin{center}
		\noindent{\Large \textbf{Capacity of entanglement and volume law}}\\
		
		\vspace*{15mm}
		\vspace*{1mm}
		\vspace*{1mm}
		{M. Reza Mohammadi Mozaffar}
		
		\vspace*{1cm}
		
		{\it Department of Physics, University of Guilan,
			P.O. Box 41335-1914, Rasht, Iran
		}
		
		\vspace*{0.5cm}
		{E-mail: {\tt mmohammadi@guilan.ac.ir}}
		
		\vspace*{1cm}
	\end{center}
	
	\begin{abstract}

We investigate various aspects of capacity of entanglement in certain setups whose entanglement entropy becomes extensive and obeys a volume law. In particular, considering geometric decomposition of the Hilbert space, we study this measure both in the vacuum state of a family of non-local scalar theories and also in the squeezed states of a local scalar theory. We also evaluate field space capacity of entanglement between interacting scalar field theories. We present both analytical and numerical evidences for the volume law scaling of this quantity in different setups and discuss how these results are consistent with the behavior of other entanglement measures including Renyi entropies. Our study reveals some generic properties of the capacity of entanglement and the corresponding reduced density matrix in the specific regimes of the parameter space. Finally, by comparing entanglement entropy and capacity of entanglement, we discuss some implications of our results on the existence of consistent holographic duals for the models in question.

\end{abstract}

\end{titlepage}
\newpage

\tableofcontents
\noindent
\hrulefill


\section{Introduction}\label{intro}

Entanglement has emerged as a topic of extensive interest in a wide variety range of
research areas ranging from quantum information theory to high energy
physics (see \cite{Eisert:2008ur,Calabrese:2009qy,Laflorencie:2015eck,Nishioka:2018khk,Casini:2022rlv} for reviews). In particular, in order to quantify quantum entanglement different measures have been studied so far including entanglement and Renyi entropies. It is well-known that for a system in a pure state, the entanglement entropy is the unique quantity which measures the amount of quantum entanglement between two complement subsystems. To define this quantity, let us assume that for a bipartite system the total Hilbert space can be
written as a direct product of two spaces $\mathcal{H}_{\rm tot}=\mathcal{H}_A\otimes \mathcal{H}_{B}$ corresponding to those of subsystems $A$ and $B$. Now summing over the degrees of freedom in $\mathcal{H}_{B}$ we find the reduced density matrix for $A$, \textit{i.e.}, $\rho_A={\rm Tr}_{B}\;\rho_{\rm tot}$. Then the entanglement entropy of the subsystem $A$ defined as the von Neumann entropy of the reduced density matrix $\rho_A$ as follows
\begin{eqnarray}\label{EE}
S_{E}=-{\rm Tr}_A\left(\rho_A \log \rho_A\right).
\end{eqnarray}
Moreover, employing the definition of the modular Hamiltonian, \textit{i.e.}, $H_A=-\log \rho_A$, one can show that $S_E=\langle H_A\rangle$. Another interesting measure which has been widely studied is the Renyi entropy defined as
\begin{eqnarray}\label{renyi}
S_{n}=\frac{1}{1-n}\log {\rm Tr}\rho_A^n,
\end{eqnarray}
where $n$ is a positive integer. This quantity contains the information of the reduced density matrix spectrum and reduces to $S_E$ in $n\rightarrow 1$ limit. Further, to gain more information about the spectrum of $\rho_A$ we can consider the capacity of entanglement which is given by \cite{Yao:2010woi}
\begin{eqnarray}\label{capa}
C_E=\lim_{n\rightarrow 1} C_n=\lim_{n\rightarrow 1} n^2\frac{\partial^2}{\partial n^2}\left((1-n)S_n\right),
\end{eqnarray}
where $C_n$ is the $n$-th capacity of entanglement. It is relatively simple to show that the above quantity gives the variance of the modular Hamiltonian, \textit{i.e.}, $C_E=\langle H_A^2 \rangle-\langle H_A \rangle^2$. In this sense, the capacity of entanglement distinguishes the width of the eigenvalue spectrum of $\rho_A$. For example, considering a maximally entangled state which has a flat entanglement spectra, the corresponding Renyi entropies are independent of $n$ and $C_E$ vanishes.

While the entanglement entropy and capacity of entanglement are intrinsically different quantities, they exactly coincide with each other in certain setups, \textit{e.g.}, quantum field theories dual to classical Einstein
gravity\cite{DeBoer:2018kvc}. An interesting question is that whether this result also hold for more general QFTs without a holographic dual. Moreover, in \cite{Arias:2023kni} the authors define a $c$-function based on the capacity of entanglement similar to the one based on $S_E$, which displays a monotonic behavior under the renormalisation group flow generated by the mass parameter. See also \cite{Nakaguchi:2016zqi,Nakagawa:2017wis,Okuyama:2021ylc,Nandy:2021hmk,Bhattacharjee:2021jff,Wei:2022bed,Shrimali:2022bvt,Andrzejewski:2023dja,Ren:2024qmx,Banks:2024cqo,Shrimali:2024nbc,MohammadiMozaffar:2024uyp} for other developments on the general properties of the capacity of entanglement in different setups both in the field theory and holography.

Let us mention that the most common way available in the literature for studying quantum
entanglement is to compute entanglement measures between two different spatial regions of a field theory. 
In this case, $A$ and $B$ are two spatial subregions at a constant time slice and the resultant measure depends on the geometric structure of the entangling surface $\partial A$ which is an artificial co-dimension-two boundary between these two subregions.\footnote{The corresponding entanglement entropy is sometimes called the geometric entropy.} In particular, entanglement entropy is UV divergent in the continuum limit of a local quantum field theory such that the coefficient of the leading term is proportional to the area of $\partial A$
\begin{eqnarray}\label{arealaw}
S_E=\mathcal{S}_{d-2}\frac{{\rm area}(\partial A)}{\epsilon^{d-2}}+\cdots,
\end{eqnarray}
where $\epsilon$ denotes the lattice spacing (the inverse of the UV cut-off) and $\mathcal{S}_{d-2}$ is a non universal (scheme dependent) constant. Note that in order to avoid the ultraviolet divergences in the continuum limit, we regulate the theory by placing it on a spatial lattice. 

The area law scaling is a consequence of the short distance correlations that exist across the entangling surface between modes which reside on different sides of $\partial A$. This specific behavior, however, does not always describe the scaling of the entanglement entropy in generic situations. For example in a $(1+1)$-dimensional critical chain, which is described by a conformal field theory in continuum limit, the leading behavior of entanglement entropy is logarithmic. Remarkably, a volume law scaling of the entanglement entropy can also be achieved in certain setups, \textit{e.g.}, non-local scalar field theories \cite{Shiba:2013jja} and lattice models with broken relativistic translational or scaling invariance \cite{Vitagliano:2010db,MohammadiMozaffar:2017nri,He:2017wla}.

Another interesting setup for which the volume law appears is dealing with entanglement in the field space, obtained by summing over a subset of the quantum fields which is meaningful when more than one field lives in a theory \cite{Taylor:2015kda}. As a typical example consider two weakly interacting scalar filed theories which live in a flat spacetime with the following action
\begin{eqnarray}\label{twofield}
I=\int d^d{x}\left(\mathcal{L}_1(\phi)+\mathcal{L}_2(\psi)+\mathcal{L}_{\rm int.}(\phi, \psi)\right),
\end{eqnarray}
where the first two terms denote the Lagrangian density of free fields and $\mathcal{L}_{\rm int.}(\phi, \psi)$ contains the interacting terms. Assuming $\mathcal{H}_{\rm tot.}=\mathcal{H}_{\phi} \otimes \mathcal{H}_{\psi}$, the corresponding reduced density matrix can be found by summing either 
$\phi$ or $\psi$, \textit{e.g.}, $\rho_{\psi}={\rm Tr}_{\phi}\left(\rho\right)$. Then the entanglement entropy is $S_E=-{\rm Tr}_{\psi}\left(\rho_{\psi}\log \rho_{\psi}\right)$. As shown in \cite{Mollabashi:2014qfa} the (field space) entanglement entropy obeys the volume law scaling behavior as expected. Moreover, $S_E$ is a monotonically increasing function of the strength of interactions between the two fields. Also considering a generalization of eq. \eqref{twofield} to field theories consisting of $N$ number of interacting scalar fields, one can study other entanglement measures including mutual information \cite{MohammadiMozaffar:2015clv}. Related investigations attempting to better understand different aspects of field space entanglement both in the field theory and holography have also appeared in \cite{Karch:2014pma,Huffel:2017ewr,Nakai:2017qos}.

Despite extensive discussions of different aspects of entanglement entropy in theories with volume law scaling, the
behavior of other entanglement measures and in particular the capacity of entanglement has not been thoroughly investigated. Therefore, in the present paper, we explore the behavior of entanglement spectrum and capacity of entanglement in such theories. To do so, we consider both geometric and field space decompositions of the Hilbert space. Moreover, to get a better understanding of the results, we will also compare the behavior of different measures, \textit{e.g.}, Renyi entropies, in various setups. We also investigate the existence of consistent holographic duals for these specific setups based on the relation between the entanglement entropy and the capacity of entanglement.

This paper is organized as follows: In section \ref{sec:palter}, we consider a simple setup which exhibits volume law scaling, the so-called $p$-alternating lattice, both in the vacuum and thermal states. By restricting the reduced subsystem to periodic sublattices, we can compute the entanglement spectrum and entanglement measures analytically. In section \ref{sec:nonlocal} we evaluate the capacity of entanglement for ground state of a family of non-local scalar field theories. Specifically, we present a combination of numerical and analytic results on the scaling of different entanglement measures in the presence of non-local correlations. In section \ref{sec:squeezed}, we extend our studies to  specific Gaussian states, the so-called squeezed states, where the corresponding entanglement entropy obeys the volume law. We investigate the capacity of entanglement for spherical entangling regions in $(3+1)$-dimensions and discuss behaviors of this quantity in different regimes. Next, we study field space capacity of entanglement in different interacting scalar field theories in section \ref{sec:fieldspace}. We conclude with a discussion of our results, as well as possible future directions, in section \ref{sec:diss}.

\section{Periodic sublattices} \label{sec:palter}
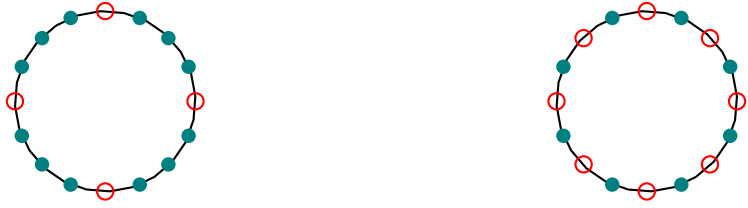
\begin{figure}
\begin{center}
\begin{tikzpicture}[scale=1.2]
\draw [thick,domain=393+90:33+90] plot ({cos(\x)}, {sin(\x)});
\draw [thick,domain=393+90:33+90] plot ({6+cos(\x)}, {sin(\x)});

\draw[red,thick] (1,0) circle (0.09);
\fill[teal] (0.9238,0.3826) circle (0.08);
\fill[teal] (0.7,0.7) circle (0.08);
\fill[teal] (0.3826,0.9238) circle (0.08);
\draw[red,thick] (0,1) circle (0.09);
\draw[red,thick] (0,-1) circle (0.09);

\draw[red,thick] (-1,0) circle (0.09);
\fill[teal] (-0.9238,0.3826) circle (0.08);
\fill[teal] (-0.7,0.7) circle (0.08);
\fill[teal] (-0.3826,0.9238) circle (0.08);

\fill[teal] (-0.9238,-0.3826) circle (0.08);
\fill[teal] (-0.7,-0.7) circle (0.08);
\fill[teal] (-0.3826,-0.9238) circle (0.08);

\fill[teal] (0.9238,-0.3826) circle (0.08);
\fill[teal] (0.7,-0.7) circle (0.08);
\fill[teal] (0.3826,-0.9238) circle (0.08);

\draw[red,thick] (6+1,0) circle (0.09);
\fill[teal] (6+0.9238,0.3826) circle (0.08);
\draw[red,thick] (6+0.7,0.7) circle (0.09);
\fill[teal] (6+0.3826,0.9238) circle (0.08);
\draw[red,thick] (6+0,1) circle (0.09);
\draw[red,thick] (6+0,-1) circle (0.09);

\draw[red,thick] (6-1,0) circle (0.09);
\fill[teal] (6-0.9238,0.3826) circle (0.08);
\draw[red,thick] (6-0.7,0.7) circle (0.09);
\fill[teal] (6-0.3826,0.9238) circle (0.08);

\fill[teal] (6-0.9238,-0.3826) circle (0.08);
\draw[red,thick] (6-0.7,-0.7) circle (0.09);
\fill[teal] (6-0.3826,-0.9238) circle (0.08);

\fill[teal] (6+0.9238,-0.3826) circle (0.08);
\draw[red,thick] (6+0.7,-0.7) circle (0.09);
\fill[teal] (6+0.3826,-0.9238) circle (0.08);
\end{tikzpicture}
\end{center}
\caption{ Examples of $p$-alternating lattice subregions for $\widetilde{N}=4, p=4$ (left) and $\widetilde{N}=8, p=2$ (right). The red open circles denote the sites within the entangling region.}
\label{fig:palter1}
\end{figure}

The first model we consider is a free scalar field theory in 1+1 dimensions whose discretized Hamiltonian is given by
\begin{eqnarray}\label{hamildis1}
H=\frac{1}{2}\sum_{i=0}^{N-1} \left(\pi_i^2+\left(\phi_{i+1}-\phi_{i}\right)^2+m^2\phi_{i}^2\right),
\end{eqnarray}
where $N$ denotes the number of total lattice sites. Here we set the lattice spacing equal to unity, \textit{i.e.}, $\epsilon=1$ and periodic boundary condition implies $\phi_{N}=\phi_0$. The calculation of different entanglement measures for this specific model have already been extensively studied in the literature, \textit{e.g.,} see \cite{Nishioka:2018khk} and references therein. In the following, we will focus on a particular entangling region which is a $p$-alternating sublattice on a lattice with periodic boundary condition, which was introduced in \cite{He:2016ohr}. These authors considered a spatial subregion consisting of $\widetilde{N}$ evenly spaced lattice sites such that $N=p\widetilde{N}$ (see figure \ref{fig:palter1}). In what follows, we will compute different entanglement measures in this setup both in the vacuum and thermal states. 

In order to find the entanglement measures we use the so-called correlator method which is an efficient algorithm to compute the eigenvalues of the reduced density matrix in general Gaussain states, \textit{e.g.}, see \cite{Peschel:2002yqj,Casini:2009sr,Eisler:2009vye}. Indeed for a thermal state we do so by first considering the following two-point functions
\begin{eqnarray}\label{correlatorXP}
X_{ij}\equiv\langle \phi_i\phi_j\rangle=\frac{1}{2N}\sum_{k=0}^{N-1}\frac{1}{\omega_k}\coth\frac{\omega_k}{2T}\cos\frac{2\pi(i-j)k}{\widetilde{N}},\nonumber\\
P_{ij}\equiv\langle \pi_i\pi_j\rangle=\frac{1}{2N}\sum_{k=0}^{N-1}\omega_k\coth\frac{\omega_k}{2T}\cos\frac{2\pi(i-j)k}{\widetilde{N}},
\end{eqnarray}
where
\begin{eqnarray}\label{dispersion}
\omega_k^2=m^2+\left(2\sin\frac{\pi k}{N}\right)^2,
\end{eqnarray}
is the corresponding dispersion relation which can be derived from eq. \eqref{hamildis1} and $i, j\in [0, \widetilde{N}-1]$. Next, defining the matrix $\sqrt{X.P}$ whose eigenvalues are $\{\lambda_j\}$, the corresponding expression for the Renyi entropy becomes
\begin{eqnarray}\label{Renyi}
S_n=\frac{1}{n-1}\sum_{j=0}^{\widetilde{N}-1}\log\left(\left(\lambda_j+\frac{1}{2}\right)^n-\left(\lambda_j-\frac{1}{2}\right)^n\right).
\end{eqnarray}
Further, using eqs. \eqref{EE} and \eqref{capa} one can also find the entanglement entropy and capacity of entanglement as follows
\begin{eqnarray}
S_E&=&\sum_{j=0}^{\widetilde{N}-1}\left(\left(\lambda_j+\frac{1}{2}\right)\log\left(\lambda_j+\frac{1}{2}\right)-\left(\lambda_j-\frac{1}{2}\right)\log\left(\lambda_j-\frac{1}{2}\right)\right),\label{SE}\\
C_E&=&\sum_{j=0}^{\widetilde{N}-1}\left(\lambda_j^2-\frac{1}{4}\right)\left(\log\frac{\lambda_j-\frac{1}{2}}{\lambda_j+\frac{1}{2}}\right)^2\label{CE}.
\end{eqnarray}
Note that using the uncertainty principle, it is a simple exercise to show that $\lambda_j\geq\frac{1}{2}$. An interesting observation is that for $p$-alternating lattice subregions the spectrum of the matrix $\sqrt{X.P}$ can be determined in terms of the corresponding eigenvalues of a circulant covariance matrix. We relegate certain details
of the calculations and refer the interested reader to \cite{He:2016ohr} for further details. It is relatively simple to show that in this set-up the eigenvalues can be written as
\begin{eqnarray}\label{lambdajplat}
\lambda_j\hspace*{-0.05cm}=\hspace*{-0.05cm}\frac{1}{4p}\bigg[\hspace*{-0.05cm}\sum_{k=0}^{p-1}\hspace*{-0.15cm}\left(\frac{\coth\frac{\omega_{k\widetilde{N}+j}}{2T}}{\omega_{k\widetilde{N}+j}}\hspace*{-0.1cm}+\hspace*{-0.1cm}\frac{\coth\frac{\omega_{(k+1)\widetilde{N}-j}}{2T}}{\omega_{(k+1)\widetilde{N}-j}}\right)\hspace*{-0.1cm}\sum_{l=0}^{p-1}\hspace*{-0.1cm}\left(\omega_{l\widetilde{N}+j}\coth\frac{\omega_{l\widetilde{N}+j}}{2T}\hspace*{-0.07cm}+\hspace*{-0.07cm}\omega_{(l+1)\widetilde{N}-j}\coth\frac{\omega_{(l+1)\widetilde{N}-j}}{2T}\right)\hspace*{-0.15cm}\bigg]^{\frac{1}{2}}\hspace*{-0.2cm}.
\end{eqnarray}
Having the corresponding eigenvalues, it is then possible to determine the behavior of the entanglement measures. To gain some intuition for the problem, in what follows, we first consider some simple cases where the spectrum has a closed form and the capacity of entanglement in various configurations can be determined analytically. Also we carry out a perturbative analysis for calculating different measures in the specific regimes of the parameter space. 

\subsection{Alternating lattice}

\begin{figure}[h!]
	\begin{center}
\includegraphics[scale=0.59]{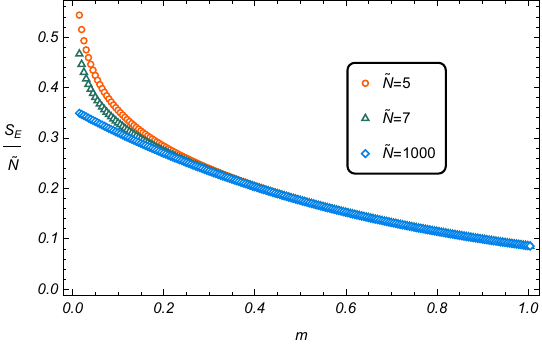}
\includegraphics[scale=0.59]{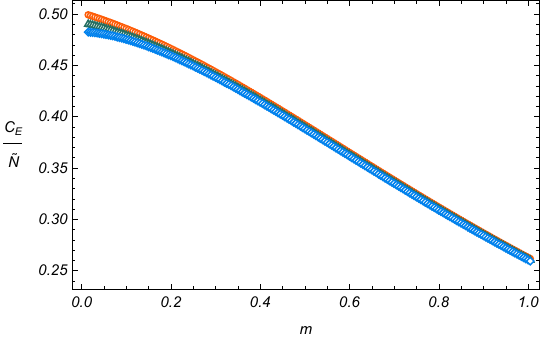}
\includegraphics[scale=0.59]{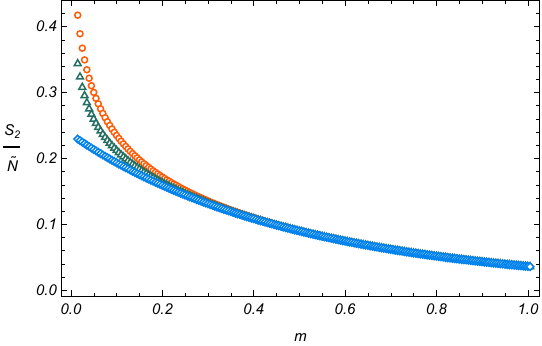}
  	\end{center}
	\caption{Entanglement entropy (left), capacity of entanglement (middle) and Renyi entropy for $n=2$ (right) as functions of mass for $p=2$ and several values of $\widetilde{N}$ at zero temperature.
	}
	\label{fig:SCSnmassp2}
\end{figure}

We begin by considering a $p=2$ alternating lattice at finite temperature. Before examining
the full behavior of the capacity of entanglement, we consider two coupled harmonic oscillators, \textit{i.e.}, $N=2$, as the simplest example. In this case, there is only one eigenvalue and eq. \eqref{lambdajplat} reduces
to
\begin{eqnarray}\label{lambdaN2p2}
\lambda=\frac{1}{4}\left(\coth^2\frac{\omega_{0}}{2T}+\coth^2\frac{\omega_{1}}{2T}+\left(\frac{\omega_0}{\omega_1}+\frac{\omega_1}{\omega_0}\right)\coth\frac{\omega_{0}}{2T}\coth\frac{\omega_{1}}{2T}\right)^{\frac{1}{2}},
\end{eqnarray}
where $\omega_0=m$ and $\omega_1=\sqrt{m^2+4}$. Specifically, in zero temperature limit we recover the vacuum state result, \textit{i.e.}, $\lambda_{\rm vac.}=\frac{\omega_0+\omega_1}{4\sqrt{\omega_0\omega_1}}$. Upon substituting
the latter expression into eqs. \eqref{Renyi}, \eqref{SE} and \eqref{CE} and expand for small mass, the resulting measures are
\begin{eqnarray}
S_E&=&\frac{1}{2}\log\frac{1}{8m}+1+\mathcal{O}(m),\label{SEN2p2}\\
S_n&=&\frac{1}{2}\log\frac{1}{8m}+\frac{\log n}{n-1}+\mathcal{O}(m),\label{SnN2p2}\\
C_E&=&1-\mathcal{O}(m)\label{CEN2p2}.
\end{eqnarray}
We see that both entanglement and Renyi entropies diverge in the massless limit which is due to the existence of a zero mode. Indeed, as shown in \cite{Calabrese:2009qy} one way to get rid of this zero mode is to break the translational symmetry of the
system, \textit{e.g.}, replacing the periodic boundary condition with the Dirichlet boundary condition. On the other hand,  the capacity of entanglement remains finite in this limit as expected. Moreover, in small mass regime it is straightforward to show that temperature corrections are exponentially suppressed, although we do not explicitly show the corresponding results here. Further, at finite temperature with vanishingly small mass we have
\begin{eqnarray}
S_E&=&\log\frac{T}{2m}+1+\frac{1}{2}\log\left(1+\frac{1}{T}\coth\frac{1}{ T}\right)+\mathcal{O}(m^2),\nonumber\\
S_n&=&\log\frac{T}{2m}+\frac{\log n}{n-1}+\frac{1}{2}\log\left(1+\frac{1}{T}\coth\frac{1}{ T}\right)+\mathcal{O}(m^2),\\
C_E&=&1-\mathcal{O}(m^2).\nonumber
\end{eqnarray}
Now let us generalize these results to a chain of harmonic oscillators on an alternating lattice. In this case we have $\widetilde{N}$ eigenvalues and eq. \eqref{lambdajplat} simplifies to
\begin{eqnarray}\label{lambdajplat3}
\lambda_j=\frac{1}{4}\bigg[\left(\frac{\coth\frac{\omega_{j}}{2T}}{\omega_{j}}+\frac{\coth\frac{\omega_{\widetilde{N}+j}}{2T}}{\omega_{\widetilde{N}+j}}\right)\left(\omega_{j}\coth\frac{\omega_{j}}{2T}+\omega_{\widetilde{N}+j}\coth\frac{\omega_{\widetilde{N}+j}}{2T}\right)\bigg]^{\frac{1}{2}}.
\end{eqnarray}
Moreover, at zero temperature the above expression becomes 
\begin{eqnarray}\label{lambdajplat3T0}
\lambda_{j, {\rm vac.}}=\frac{\omega_{j}+\omega_{\widetilde{N}+j}}{4\sqrt{\omega_{j}\omega_{\widetilde{N}+j}}}.
\end{eqnarray}
Using the above result we present the mass dependence of various measures at zero temperature for several values of $\widetilde{N}$ in figure \ref{fig:SCSnmassp2}. Note that for evaluating the Renyi entropy we focus on $n=2$, because the interesting qualitative features of this measure are independent of the Renyi index. We find a number of key features:
First, all the measures are extensive for sufficiently large values of $\widetilde{N}$ and $m$. In particular, $C_E/\widetilde{N}$ becomes constant in this limit which is reminiscent of a volume law scaling. Second, although the entanglement and Renyi entropies diverge in the massless limit, the capacity of entanglement saturates to a finite value in agreement with eqs. \eqref{SEN2p2} and \eqref{CEN2p2}. Further, the Renyi entropy is a decreasing function of the Renyi index as expected. Moreover, all the measures are monotonically decreasing as we increase the mass parameter.  
Remarkably, in the massless case when we consider large $N$ limit one can find the leading scaling of the measures for the vacuum state analytically. In \cite{He:2016ohr} the authors found the leading behavior of the entanglement entropy using this approach. Indeed, in the limit of large $N$, neglecting the zero mode contribution we can express the measures as an integral over a continuous variable defined as $x\equiv j/N$
\begin{eqnarray}\label{largeNplat}
S_n&=&\frac{2\widetilde{N}}{n-1}\int_{0}^{\frac{1}{2}}dx\log\left(\left(\lambda(x)+\frac{1}{2}\right)^n-\left(\lambda(x)-\frac{1}{2}\right)^n\right),\nonumber\\
C_E&=&2\widetilde{N}\int_{0}^{\frac{1}{2}}dx\left(\lambda^2(x)-\frac{1}{4}\right)\left(\log\frac{\lambda(x)-\frac{1}{2}}{\lambda(x)+\frac{1}{2}}\right)^2,
\end{eqnarray}
where $\lambda(x)$ can be determined by simply evaluating eqs. \eqref{dispersion} and \eqref{lambdajplat3T0} in this limit, \textit{i.e.}, 
\begin{eqnarray}\label{lambdajplat3T1}
\lambda(x)=\frac{\sin\pi x+\cos\pi x}{4\sqrt{\sin\pi x\;\cos\pi x}}.
\end{eqnarray}
Also the entanglement entropy has a similar expression which can be found using eq. \eqref{SE}. Evaluating numerically the integrals for the entanglement entropy and capacity of entanglement yield
\begin{eqnarray}
S_E=0.36 \;\widetilde{N},\hspace*{2cm}C_E=0.48 \;\widetilde{N},
\end{eqnarray}
and thus both measures obey a volume law and become extensive. Note that the numerical coefficients are consistent with the numerical results reported in figure \ref{fig:SCSnmassp2} in the large $N$ limit and small mass regime. Further, the Renyi entropy for some values of the Renyi index is given by
\begin{eqnarray}
S_2=\left(\frac{2C}{\pi}-\log\sqrt{2}\right)\;\widetilde{N} \sim 0.24\; \widetilde{N},\hspace*{2cm}S_3=0.2 \;\widetilde{N},
\end{eqnarray}
where $C$ denotes the Catalan's constant. Again we see that the Renyi entropies become extensive and the qualitative behaviors are consistent with figure \ref{fig:SCSnmassp2}.

To close this subsection, let us comment on extending this discussion to finite temperature case. Indeed, in this case our numerical results show that all the aforementioned measures increase with the temperature and further the qualitative dependence on mass is similar to the corresponding results for the vacuum state. Moreover, in the high temperature limit we can find the leading behavior of these quantities rather easily. In fact, in this limit eq. \eqref{lambdajplat3} can be approximated by
\begin{eqnarray}
\lambda_j\sim\frac{T}{\sqrt{2}}\left(\frac{1}{\omega_{j}^2}+\frac{1}{\omega_{j+N_A}^2}\right)^{\frac{1}{2}}.
\end{eqnarray}
Noting that in $1\ll m\ll T$ limit the eigenfrequencies become equal (see eq. \eqref{dispersion}) and the above expression then simplifies to $\lambda_j\sim \frac{T}{m}$. 
In this case the corresponding expressions for the measures at leading order become 
\begin{eqnarray}
S_E\sim S_n\sim \widetilde{N}\log \frac{T}{m},\hspace*{2.5cm}C_E\sim \widetilde{N}.
\end{eqnarray}
Indeed, based on the above results we see that at leading order
all the measures are proportional to the volume. Interestingly, we see that in this limit $C_E\ll S_E$ which shows that the reduced density matrix becomes more and more maximally mixed as one increases the temperature as expected.

\subsection{$p$-alternating lattice and the continuum limit}
\begin{figure}[h!]
	\begin{center}
\includegraphics[scale=0.59]{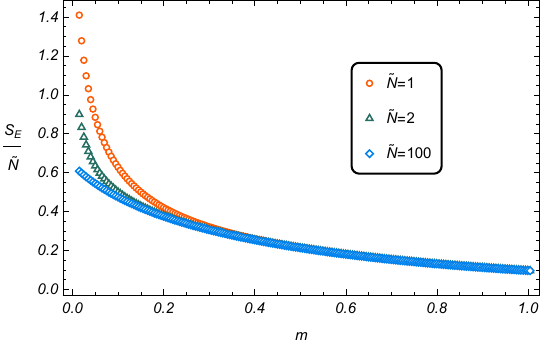}
\includegraphics[scale=0.59]{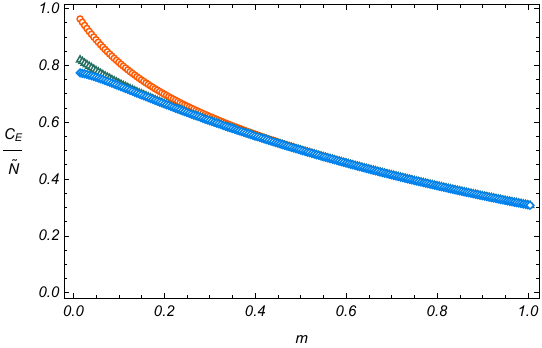}
\includegraphics[scale=0.59]{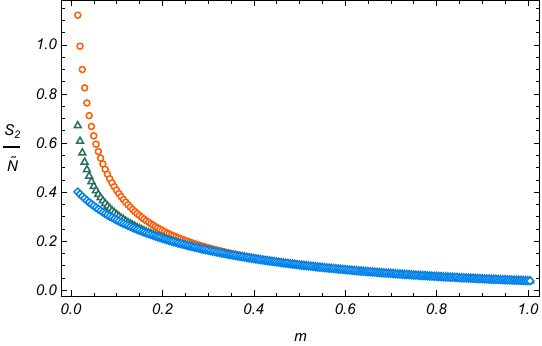}
  	\end{center}
	\caption{Entanglement entropy (left), capacity of entanglement (middle) and Renyi entropy for $n=2$ (right) as functions of mass for $p=10$ and several values of $\widetilde{N}$.
	}
	\label{fig:SCSnmassp10}
\end{figure}

In this section we generalize our studies to a $p$-alternating lattice of coupled harmonic oscillators, where the corresponding eigenvalues are given by eq. \eqref{lambdajplat}, in specific directions. We will focus on the zero temperature limit corresponds to the vacuum state. Some numerical results for $p=10$ are illustrated in figure \ref{fig:SCSnmassp10}. In comparing these results with figure \ref{fig:SCSnmassp2}, we see that the entanglement measures increase for larger values of $p$ such that their qualitative behavior does not depend strongly on this parameter.

\begin{figure}[h!]
	\begin{center}
\includegraphics[scale=0.8]{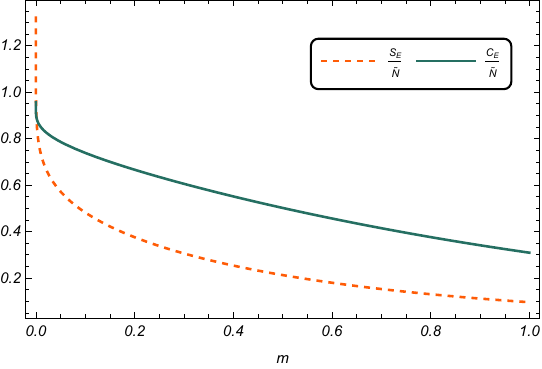}
\hspace*{1cm}
\includegraphics[scale=0.8]{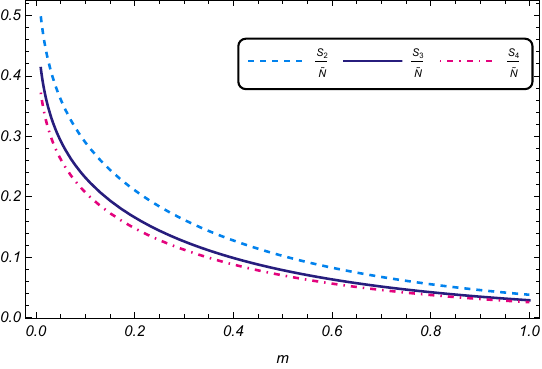}
  	\end{center}
	\caption{Entanglement entropy and capacity of entanglement (left) and Renyi entropy (right) as functions of mass in $p\rightarrow \infty$ limit.}
	\label{fig:SCSnmasspinf}
\end{figure}

Remarkably, we can find the behavior of the entanglement measures in $p\gg 1$ limit for fixed value of $\widetilde{N}$. As shown in \cite{He:2016ohr} in this case the spectrum becomes degenerate and the corresponding expression for the entanglement entropy becomes
\begin{eqnarray}
S_E=\widetilde{N}\left(\left(\lambda+\frac{1}{2}\right)\log\left(\lambda+\frac{1}{2}\right)-\left(\lambda-\frac{1}{2}\right)\log\left(\lambda-\frac{1}{2}\right)\right),
\end{eqnarray}
with $\lambda=\frac{1}{\pi}\sqrt{K\left(-4/m^2\right)E\left(-4/m^2\right)}$ where $K(x)$ and $E(x)$ denote the elliptic integrals. Similarly for the capacity of entanglement using eq. \eqref{CE} we have
\begin{eqnarray}
C_E=\widetilde{N}\left(\lambda^2-\frac{1}{4}\right)\left(\log\frac{\lambda-\frac{1}{2}}{\lambda+\frac{1}{2}}\right)^2.
\end{eqnarray}
Indeed, in both cases the volume law scaling is clearly manifest. Figure \ref{fig:SCSnmasspinf} illustrates the behavior of the above quantities as functions of the mass parameter. For completeness, we also plot the Renyi entropy for several values of $n$. Again, all the measures are monotonically decreasing as we increase $m$. Indeed, we can make use of the above expressions to find the small mass expansion of the measures as follows 
\begin{eqnarray}
S_E=\frac{\widetilde{N}}{2}\log\left(\log\frac{1}{m}\right)+\mathcal{O}(1), \hspace*{2cm}C_E=\widetilde{N}-\mathcal{O}\left(\frac{1}{\log m}\right).
\end{eqnarray}
Interestingly, we see that in this case the divergence of the entanglement entropy is double logarithm which is much softer than the case with finite $p$ (see eq. \eqref{SEN2p2}). Moreover, in this limit $C_E/S_E\ll 1$ which shows that the reduced density matrix becomes more and more maximally mixed as one decreases the mass parameter. A simple analysis shows that these results hold qualitatively even in a finite temperature state.

\section{Non-local scalar theories} \label{sec:nonlocal}
In this section we evaluate capacity of entanglement for a family of non-local scalar field theories with the following Hamiltonian in $1+1$ dimensions \cite{Shiba:2013jja}
\begin{eqnarray}\label{shibataka}
H=\frac{1}{2}\int dx \left(\left(\partial_t\phi\right)^2+\phi e^{A(-\partial_x\partial^x)^{w/2}}\phi\right),
\end{eqnarray}
where $A$ and $w$ are constants. Note that due the presence of spatial higher derivatives the number of lattice sites which are correlated together depends on $w$. Also by increasing $A$ the correlation strength between the lattice sites becomes more pronounced. Indeed, we will see that different choices of these parameters will modify leading contributions to entanglement measures in the following. Moreover, the Hamiltonian is still second order in time derivative and thus the corresponding conjugate momentum is $\pi=\partial_t\phi$ as expected.

Again, we employ the correlator method and use eqs. \eqref{SE} and \eqref{CE} to find the corresponding entanglement measures. However, unlike in the Klein-Gordon case, which is a two derivative theory, now the higher derivative terms on a discrete lattice are more involved. It is relatively straightforward to diagonalize $(-\partial_x\partial^x)^{w/2}$ by a Fourier transform, which leads to the following relations for the two point functions
\begin{eqnarray}\label{XP}
X_{mn}&=&\frac{1}{2\pi}\int_{-\pi}^{\pi}e^{-\frac{A}{2}\left(2\sin\frac{k}{2}\right)^w}e^{ik(m-n)}dk,\\
P_{mn}&=&\frac{1}{2\pi}\int_{-\pi}^{\pi}e^{\frac{A}{2}\left(2\sin\frac{k}{2}\right)^w}e^{ik(m-n)}dk.
\end{eqnarray}

\begin{figure}[h!]
	\begin{center}
	\includegraphics[scale=0.85]{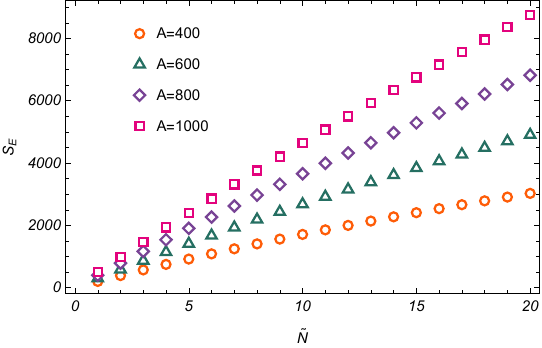}
  \hspace*{0.6cm}
\includegraphics[scale=0.85]{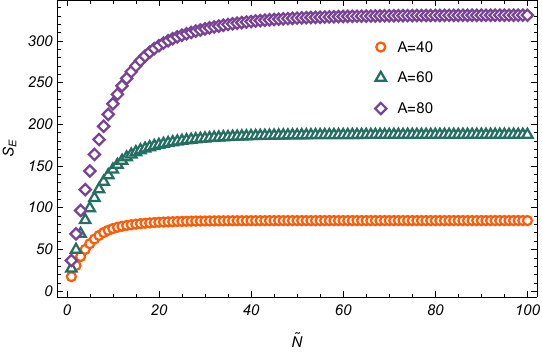}
  	\end{center}
	\caption{Entanglement entropy in $\widetilde{N}\ll A$ (left) and $\widetilde{N}\gg A$ (right) regimes for $w=1$.}
	\label{fig:13111643fig1SA}
\end{figure}

In \cite{Shiba:2013jja} it was shown that due to the non-local structure of the Hamiltonian the corresponding entanglement entropy for this model obeys a volume law.\footnote{Several aspects of quantum information theoretic measures has been studied in this non-local model, \textit{e.g.}, see \cite{Vasli:2023syq,Pirmoradian:2023uvt}.} These authors argued that this  scaling behavior happens provided that the size of subregion is smaller than a certain length scale which depends on $A$. Indeed, for specific values of $w$ the scaling of the entanglement entropy can be found numerically as follows
\begin{eqnarray}\label{nonlocalSE}
S_E\approx\Bigg\{ \begin{array}{rcl}
&c_1 A\widetilde{N}&\,\,\,\; \widetilde{N}\ll A\label{SAellllA}\\
&c_2 A^2&\,\,\,\; \widetilde{N}\gg A
\end{array},
\end{eqnarray}
where $c_1$ and $c_2$ are $w$-dependent constants and the best fit gives $c_1\approx \frac{w}{2}$. In figure \ref{fig:13111643fig1SA} we reproduce the results first found in \cite{Shiba:2013jja} for several values of $A$ with $w=1$. The numerical results depicted in the left panel show that entanglement entropy has a linear dependence on $\widetilde{N}$ in $\widetilde{N}\ll A$ limit and thus scales with the volume (instead of the area) of the entangling region. On the other hand, the right panel shows that for $\widetilde{N}\gg A$ the entanglement entropy first increases very sharply with $\widetilde{N}$ and then suddenly saturates to a constant value. Moreover, the saturation value is a monotonically increasing function of $A$ due to the enhancement of the correlation between the lattice sites. It is worth to mention that considering other values of $w$ the qualitative features of these results do not change.

\begin{figure}[h!]
	\begin{center}
\includegraphics[scale=0.86]{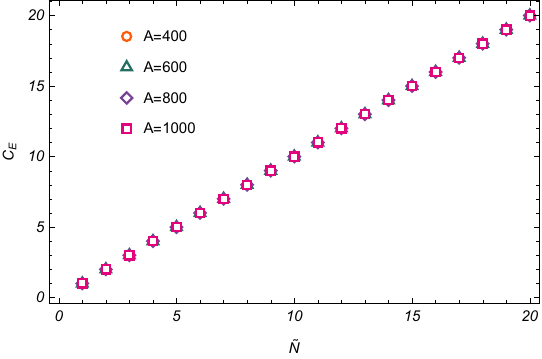}
  \hspace*{0.4cm}
\includegraphics[scale=0.86]{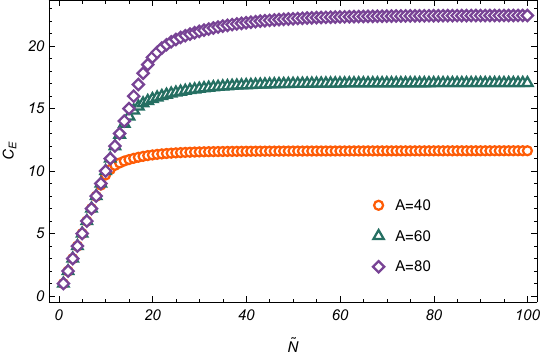}
  	\end{center}
	\caption{Capacity of entanglement in $\widetilde{N}\ll A$ (left) and $\widetilde{N}\gg A$ (right) regimes for $w=1$.}
	\label{fig:13111643fig1CA}
\end{figure}

\begin{figure}[h!]
	\begin{center}
\includegraphics[scale=0.86]{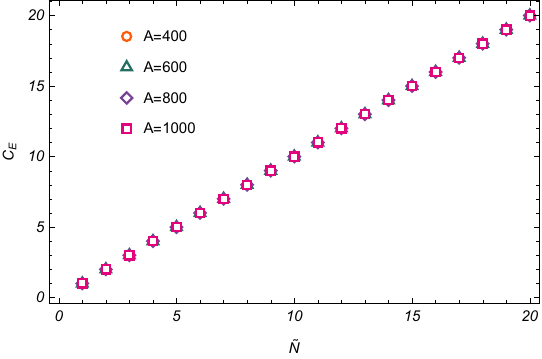}
  \hspace*{0.4cm}
\includegraphics[scale=0.86]{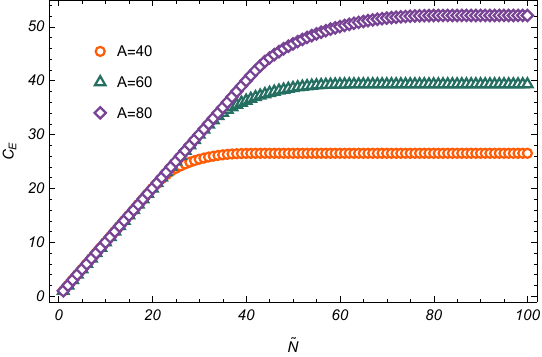}
  	\end{center}
	\caption{Capacity of entanglement in $\widetilde{N}\ll A$ (left) and $\widetilde{N}\gg A$ (right) regimes for $w=2$.
	}
	\label{fig:13111643fig2CA}
\end{figure}

Let us now turn to the computation of capacity of entanglement in this non-local model using eq. \eqref{CE}. The corresponding numerical results are summarized in figures \ref{fig:13111643fig1CA} and \ref{fig:13111643fig2CA} for $w=1, 2$ respectively. Note that we will mainly consider the same values of $A$ as in figure \ref{fig:13111643fig1SA} since this choice facilitates a comparison to the analogous results for the entanglement entropy. Based on the left panels, we observe that the capacity of entanglement grows linearly with the subregion width in $\widetilde{N}\ll A$ limit. Remarkably, comparing figures \ref{fig:13111643fig1CA} and \ref{fig:13111643fig2CA}, we see that the rate of growth of $C_E$ is independent of both $w$ and $A$. On the other hand, the right panels show that for $\widetilde{N}\gg A$ the capacity of entanglement first increases linearly with $\widetilde{N}$ and then suddenly saturates to a constant value. Further, the saturation value monotonically increases for larger values of $w$ and $A$, as expected.
Our numerical results suggest that the capacity of entanglement behaves like
\begin{eqnarray}\label{nonlocalCE}
C_E\sim \Bigg\{ \begin{array}{rcl}
&\widetilde{N}&\,\,\,\; \widetilde{N}\ll A\label{CAellllA},\\
&c_3A&\,\,\,\; \widetilde{N}\gg A,
\end{array}
\end{eqnarray}
where $c_3\approx \frac{w}{3}$. We conclude that for $\widetilde{N}\ll A$ the capacity of entanglement obeys a volume law whose coefficient is completely independent of the chosen parameters which is unexpected. Interestingly, comparing the above result with eq. \eqref{nonlocalSE} we see that in both regimes $\frac{C_E}{S_E}\sim A^{-1}$. Hence for large values of $A$ we have $C_E\ll S_E$ which is consistent with the idea that the corresponding reduced density matrix becomes more and more maximally mixed as one increases the strength of the correlation between the lattice sites.

Before we proceed further, we would like to study the asymptotic behaviors of the entanglement measures in $\widetilde{N}\ll A$ limit employing a semi analytic treatment. Indeed, this study plays an important role in our analysis in the following. Based on our numerical results the main contribution to the measures comes from the largest eigenvalue of $\sqrt{X.P}$, \textit{i.e.}, $\nu_{\rm max}$\footnote{Indeed, based on our numerics we see that other eigenvalues are less than $\mathcal{O}(10^{-3})$ times $\nu_{\rm max}$.}. Thus using eqs. \eqref{SE} and \eqref{CE} the asymptotic expansions of $S_E$ and $C_E$ for $\nu_{\rm max}\gg 1$ are
\begin{eqnarray}\label{SACAlargenu}
S_E&\sim&\sum_{k=1}^{\widetilde{N}}\left(\log \nu_{\rm max}+\mathcal{O}(1)\right)=\widetilde{N} \log \nu_{\rm max}+\cdots,\nonumber\\
C_E&\sim&\sum_{k=1}^{\widetilde{N}}\left(1+\mathcal{O}(\nu_{\rm max}^{-2})\right)=\widetilde{N}+\cdots.
\end{eqnarray}
which are consistent with eqs. \eqref{SAellllA} and \eqref{CAellllA}. Now combining the above two equations yields the following
\begin{eqnarray}
\frac{C_E}{S_E}\sim\frac{2\widetilde{N}}{\widetilde{N} \log \nu_{\rm max}}\sim\frac{1}{\log \nu_{\rm max}}\ll 1\;\rightarrow\;C_E\ll S_E,
\end{eqnarray}
which shows that the corresponding state becomes maximally mixed. Moreover, we can find the leading behavior of $\nu_{\rm max}$ in this limit to obtain a more rigorous estimation. Although this behavior is independent of $w$, let us for simplicity restrict our analysis to a specific case with $w=2$. In this case eq. \eqref{XP} yields
\begin{eqnarray}
&&X_{mn}=\frac{1}{\pi}\int_{0}^{\pi}e^{-2A\left(\sin\frac{k}{2}\right)^2}\cos\left((m-n)k\right)=e^{-A}(-i)^{m-n}J_{m-n}(iA),\nonumber\\
&&P_{mn}=\frac{1}{\pi}\int_{0}^{\pi}e^{2A\left(\sin\frac{k}{2}\right)^2}\cos\left((m-n)k\right)=e^{A}i^{m-n}J_{m-n}(iA),
\end{eqnarray}
and hence
\begin{eqnarray}
(X.P)_{mn}=\sum_{j=1}^{\widetilde{N}}J_{m-j}(iA)J_{j-n}(iA).
\end{eqnarray}
Now in $\widetilde{N}\ll A$ limit using the asymptotic behavior of the Bessel function $J_{\alpha}(z)\xrightarrow{z\gg \alpha}\sqrt{\frac{2}{\pi z}}\cos\left(z-\frac{\alpha\pi}{2}-\frac{\pi}{4}\right)$
we have
\begin{eqnarray}
(X.P)_{mn}&=&\frac{2}{\pi iA}\sum_{j=1}^{\widetilde{N}}\cos\left(iA-\frac{(m-j)\pi}{2}-\frac{\pi}{4}\right)\cos\left(iA-\frac{(j-n)\pi}{2}-\frac{\pi}{4}\right)\nonumber\\
&=&\frac{1}{2\pi iA}\left(2i\widetilde{N} \sinh\left(2A+\frac{i(m-n)\pi}{2}\right)+\sin\left(\frac{(m+n+1-2\widetilde{N})\pi}{2}\right)-\cos\left(\frac{(m+n)\pi}{2}\right)\right)\nonumber\\
&\sim&\frac{\widetilde{N}}{2\pi A}e^{2A}+\cdots\approx \nu_{\rm max}^2,
\end{eqnarray}
where in the last step we neglect the higher order terms. Plugging the above result back into eq. \eqref{SACAlargenu} we are left with
\begin{eqnarray}
S_E&\sim &\widetilde{N}A-\frac{\widetilde{N}}{2}\log A+\cdots,\\
C_E&\sim &\widetilde{N}-\frac{\pi A}{6}\;e^{-2A}+\cdots,
\end{eqnarray}
which is precisely matches with eqs. \eqref{nonlocalSE} and \eqref{nonlocalCE} for $w=2$ at leading order. 
Interestingly enough, based on these results we see that in $A \gg \widetilde{N}$ limit the relationship $C_E=S_E$
is completely broken. Generally, the entanglement entropy and the capacity of entanglement have no reason to be equal. However, it was argued in \cite{DeBoer:2018kvc} that perhaps such a relation in QFTs is a hint of a dual gravitational picture. Hence, at least in this regime our results suggest that the corresponding vacuum state may not have a solution of a classical gravity theory as a holographic dual. 

\begin{figure}[h!]
	\begin{center}
\includegraphics[scale=0.86]{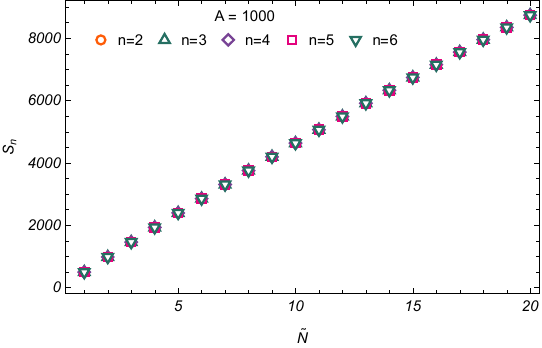}
  \hspace*{0.4cm}
\includegraphics[scale=0.86]{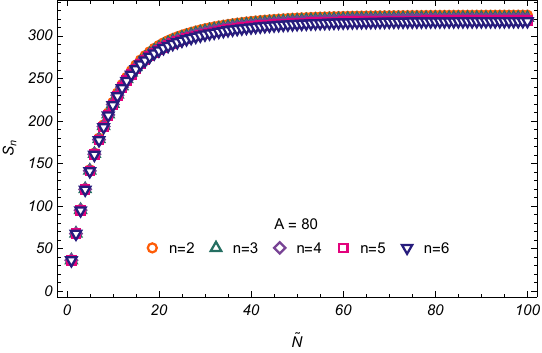}
  	\end{center}
	\caption{Renyi entropy in $\widetilde{N}\ll A$ (left) and $\widetilde{N}\gg A$ (right) regimes for several values of $n$ with $w=1$.
	}
	\label{fig:13111643fig1Sn}
\end{figure}

To close this section, let us comment on the behavior of Renyi entropy in this non-local model. We present the $n$-dependence of the Renyi entropy for several values of the parameters in figure \ref{fig:13111643fig1Sn}. From these plots, one can infer that the qualitative features of the Renyi entropy are similar to the entanglement entropy. Interestingly, from the left panel we see that in the regime where the volume law dominates, the Renyi entropy is completely independent of the Renyi index. On the other hand, the right panel shows that for $\widetilde{N}\gg A$, Renyi entropy is a decreasing function of $n$ such that the rate of change of $S_n$ is a monotonically decreasing function of the Renyi index.  

\section{Scalar fields in squeezed states} \label{sec:squeezed}

In this section we extend our studies to specific Gaussian states, the so-called squeezed states, where the corresponding entanglement entropy obeys the volume law. This intriguing behavior first observed for a $(1+1)$-dimensional scalar field theory in \cite{Katsinis:2023hqn} and then extended to higher dimensional spherical entangling regions in \cite{Katsinis:2024sek}. Of course, in our previous work we have studied several aspects of capacity of entanglement in a two dimensional setup \cite{MohammadiMozaffar:2024uyp}. Our main objective in the following is to extend this study to 3+1 dimensions with a spherical entangling surface. This is easily accomplished by generalizing the ground state calculations first performed in \cite{Bombelli:1986rw,Srednicki:1993im} to squeezed states where the corresponding wave function reads
\begin{equation}\label{squstate}
\psi(x, t)=\left(\frac{m{\Re}(w(t))}{\pi}\right)^{1/4}\exp\left(-\frac{mw(t)}{2}(x-x_0(t))^2+ip_0(t)(x-x_0(t))-i\phi_s(t)\right),
\end{equation}
where
\begin{eqnarray}\label{squstate1}
w(t)=\omega\frac{1-i\sinh z\;\cos 2\omega t}{\cosh z+\sinh z\;\sin 2\omega t},\hspace*{1cm}
\phi_s(t)=\phi_c(t)-\frac{\omega t}{2}+\frac{1}{2}\tan^{-1}\frac{\tanh\frac{z}{2}+\tan \omega t}{1+\tanh\frac{z}{2}\tan \omega t}.
\end{eqnarray}
In what follows, we consider a free massive scalar field with Hamiltonian 
\begin{eqnarray}\label{scalarhamil}
H=\frac{1}{2}\int d^3x\left(\pi^2(\vec{x})+\vec{\nabla}\phi(\vec{x}).\vec{\nabla}\phi(\vec{x})+\mu^2\phi^2(\vec{x})\right).
\end{eqnarray}
Expanding the scalar field in terms of the spherical harmonics $\phi=\sum_{\ell, m} \frac{\phi_{\ell, m}(r)Y_{\ell, m}(\theta, \varphi)}{r}$, we have
\begin{eqnarray}\label{scalarhami2}
H=\frac{1}{2}\sum_{\ell, m}\int_0^{\infty} dr\left[\pi_{\ell m}^2(r)+r^2\left(\partial_r\left(\frac{\phi_{\ell m}(r)}{r}\right)\right)^2+\left(\frac{\ell(\ell+1)}{r^2}+\mu^2\right)\phi_{\ell m}^2(r)\right],
\end{eqnarray}
where $\ell>0$ and $-\ell\leq m\leq \ell$. Further, replacing the space continuum with a discrete mesh of lattice points the discretized version of the above expression becomes
\begin{eqnarray}\label{scalarhami3}
H=\frac{1}{2\epsilon}\sum_{\ell, m}\sum_{j=1}^{N}\left[\pi_{\ell m, j}^2+(j+\frac{1}{2})^2\left(\frac{\phi_{\ell m, j}}{j}-\frac{\phi_{\ell m, j+1}}{j+1}\right)^2+\left(\frac{\ell(\ell+1)}{j^2}+\mu^2\epsilon^2\right)\phi_{\ell m, j}^2\right],
\end{eqnarray}
where $N$ denotes the number of lattice sites inside in a spherical box of radius $L=N\epsilon$. Again, without loss of generality we set the lattice spacing equal to unity, \textit{i.e.}, $\epsilon=1$. Defining 
\begin{eqnarray}\label{Kij}
K_{ij}=\left(\left(i+\frac{1}{2}\right)^2+\left(i-\frac{1}{2}\right)^2\left(1-\delta_{i,1}\right)+\ell(\ell+1)+\mu^2 i^2\right)\frac{\delta_{i,j}}{i^2}-\frac{\left(i+\frac{1}{2}\right)^2\delta_{i+1,j}}{ij}-\frac{\left(j+\frac{1}{2}\right)^2\delta_{i,j+1}}{ij},
\end{eqnarray}
we see that eq. \eqref{scalarhami3} can be rewritten as follows 
\begin{eqnarray}\label{Nhamil}
H=\sum_{\ell, m}\left(\frac{1}{2}\sum_{i=1}^N\pi_{\ell m, i}^2+\frac{1}{2}\sum_{i, j=1}^N \phi_{\ell m, i} K_{ij}\phi_{\ell m, j}\right)\equiv \sum_{\ell, m}H_{\ell, m},
\end{eqnarray}
where $H_{\ell, m}$ is the same as a Hamiltonian for a system of $N$ harmonic oscillators with specific couplings. In a similar manner to the analysis of the ground state in \cite{Srednicki:1993im}, we can diagonalize the matrix $K$ using a similarity transformation, \textit{i.e.,} $K_D=UKU^T$. Further, to evaluate the spectrum of the reduced density matrix by tracing over the first $\widetilde{N}$ oscillators one can consider the following decomposition for the square root of $K$ 
\begin{eqnarray}\label{OmegaKD}
\Omega=U^TK_D^{1/2}U\equiv\left(\begin{matrix}
A & B\\
B^T & C
\end{matrix}\right),
\hspace*{2cm}(K_{D})_{ij}=w_i\;\delta_{ij},
\end{eqnarray}
where $A$ is an $\widetilde{N}\times \widetilde{N}$ matrix and $C$ is an $(N-\widetilde{N})\times (N-\widetilde{N})$ matrix. Note that for squeezed states $w_i$'s are given by the application of eq. \eqref{squstate1} for each
normal mode and thus $\Omega$ is a complex symmetric matrix. Finally, it was shown in \cite{Katsinis:2023hqn} that in this case the entanglement entropy is given by eq. \eqref{SE} by substituting $\lambda_j\rightarrow \tilde{\xi}_j/2$ where $\tilde{\xi}_j$'s are the corresponding eigenvalues of a matrix $\tilde{\Omega}$ defined by
\begin{eqnarray}\label{tildeOmegad}
\tilde{\Omega}={\rm Re}(\Omega)^{-1}\left(\begin{matrix}
-{\rm Re}(A) & i{\rm Im}(B)\\
-i{\rm Im}(B)^T & {\rm Re}(C)
\end{matrix}\right).
\end{eqnarray}
Similarly, the capacity of entanglement can be fund by using eq. \eqref{CE}. Before we proceed, let us recall that in this setup due to the spherical symmetry the wave function of the total Hamiltonian is a direct product of the corresponding wave functions of each $H_{\ell, m}$ and hence the entanglement measures are found by summing over $\ell$ and $m$. Moreover, from eq. \eqref{Nhamil} we see that the Hamiltonian for each $\ell$-sector is independent of $m$ and thus summing over $m$ gives a factor of $2\ell+1$, \textit{i.e.},
\begin{eqnarray}\label{measuresqueez}
S_n=\sum_{\ell=0}^{\infty}(2\ell+1)S_{n,\ell},\hspace*{2cm}C_E=\sum_{\ell=0}^{\infty}(2\ell+1)C_{E,\ell}
\end{eqnarray}
Let us recall that as shown in \cite{Srednicki:1993im} the above series for the entanglement entropy with $z=0$ is convergent. Indeed, in this case it is easy to show that the value of $S_{E,\ell}$ decreases with $\ell$ such that for $\ell\gg N$ we have
\begin{eqnarray}\label{SElargeL}
S_{E,\ell}\sim \frac{\widetilde{N}(\widetilde{N}+1)(2\widetilde{N}+1)^2}{16\ell^4}\log \ell+\cdots.
\end{eqnarray}
Similarly for the capacity of entanglement the contribution of each large angular momentum sector can be computed as follows
\begin{eqnarray}\label{CElargeL}
C_{E,\ell}\sim \frac{\widetilde{N}(\widetilde{N}+1)(2\widetilde{N}+1)^2}{4\ell^4}\left(\log \ell\right)^2+\cdots.
\end{eqnarray}
From the numerical results, one can infer that for a non trivial squeezing the large angular momentum sectors exhibit qualitatively similar behavior. Hence, in what follows we consider an angular cutoff $500\leq\ell_{\rm max}\leq 1000$ when performing the numerical calculations which is perfectly consistent with the previous results reported in the literature, \textit{e.g.}, \cite{Srednicki:1993im,Lohmayer:2009sq,Kim:2014nza}. Further, based on eq. \eqref{squstate1} the spectrum
of the reduced density matrix depends on time and hence the measures are also time-dependent. To gain more insights into the behavior of the measures, we compute the mean quantities by sampling over 100 random time instances.

\begin{figure}[h!]
	\begin{center}
\includegraphics[scale=0.59]{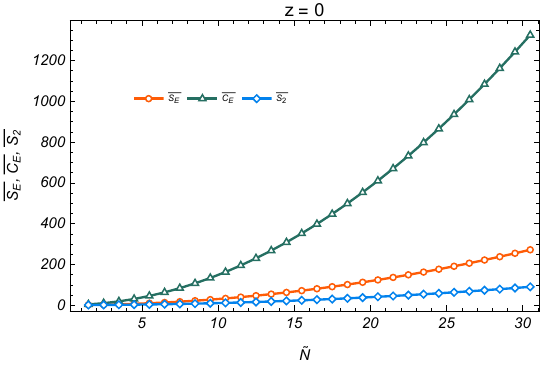}
\includegraphics[scale=0.59]{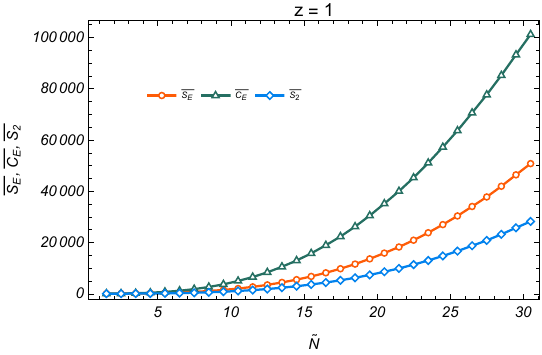}
\includegraphics[scale=0.59]{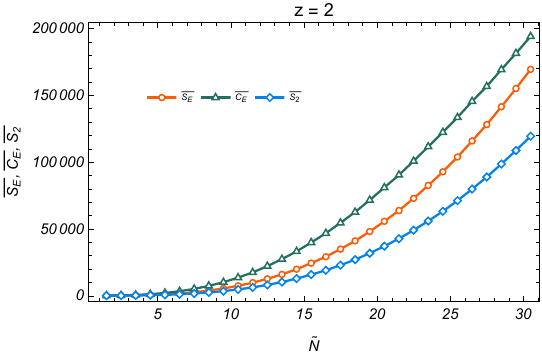}
  	\end{center}
	\caption{Mean values of the entanglement entropy, capacity of entanglement and second Renyi entropy as functions of $\widetilde{N}$ for several values of $z$ when all modes lie in a squeezed state with the same squeezing parameter. The left panel corresponds to the vacuum state with $z=0$.	}
	\label{fig:seces2z}
\end{figure}

\begin{figure}[h!]
	\begin{center}
\includegraphics[scale=0.84]{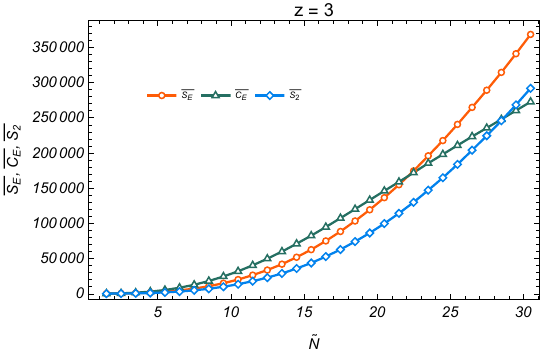}
\hspace*{0.5cm}
\includegraphics[scale=0.84]{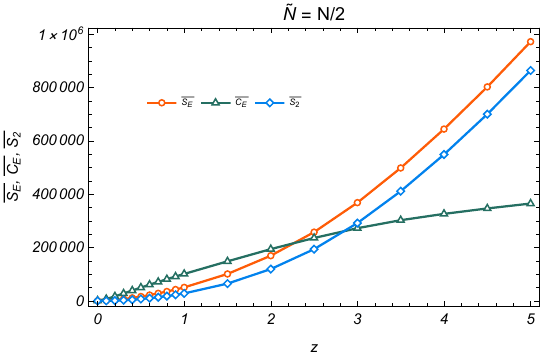}
  	\end{center}
	\caption{\textit{Left}: Mean values of the entanglement entropy, capacity of entanglement and second Renyi entropy as functions of $\widetilde{N}$ for $z=3$ in the massless limit. \textit{Right}: The same quantities as functions of the squeezing parameter for $\widetilde{N}=N/2$.	}
	\label{fig:seces2zhalf}
\end{figure}

The corresponding numerical results for different quantities in the massless limit are summarized in figures \ref{fig:seces2z} and
\ref{fig:seces2zhalf}. Here we will mainly consider $N = 60$, because this choice facilitates a comparison to the analogous results for entanglement entropy in \cite{Srednicki:1993im} for $z=0$ and \cite{Katsinis:2024sek} for larger values of the squeezing parameter.

Figure \ref{fig:seces2z} illustrates the mean values of the entanglement entropy, capacity of entanglement and second Renyi entropy as functions of $\widetilde{N}$ for several values of the squeezing parameter when all modes lie in a squeezed state with the same squeezing parameter.. The left panel corresponds to $z=0$ case where all these measures obey the area law scaling. Indeed, as can be seen, the points
are well-fitted by
\begin{eqnarray}\label{sesncefitz0}
S_{E}\sim 0.29\frac{R^2}{\epsilon^2}+\cdots,\hspace*{1cm}C_{E}\sim 1.43\frac{R^2}{\epsilon^2}+\cdots,\hspace*{1cm}S_{2}\sim 0.096\frac{R^2}{\epsilon^2}+\cdots,
\end{eqnarray}
where $R\equiv \left(\widetilde{N}+\frac{1}{2}\right)\epsilon$ denotes the radius of the spherical entangling region. Note that the coefficients of the leading area term are nonuniversal (scheme-dependent) numbers. Moreover, the middle and right panels in this figure correspond to $z=1$ and $z=2$ cases respectively. Interestingly, in these cases the leading behavior of the measures  is proportional to the volume of the sphere, \textit{e.g.}, for $z=2$ we have\footnote{Note that a single volume term fits almost perfectly the numerical data, although the combination of area and volume is more precise.} 
\begin{eqnarray}\label{sesncefitz1}
\overline{S_{E}}\sim 5.1\frac{R^3}{\epsilon^3}+29\frac{R^2}{\epsilon^2}+\cdots,\hspace*{0.6cm}\overline{C_{E}}\sim 1.8\frac{R^3}{\epsilon^3}+161\frac{R^2}{\epsilon^2}+\cdots,\hspace*{0.6cm}\overline{S_{2}}\sim 4.1\frac{R^3}{\epsilon^3}+4.8\frac{R^2}{\epsilon^2}+\cdots.
\end{eqnarray}
Indeed, our numerical results make it clear that even for small values of the squeezing parameter, the volume law scaling holds. Further, all the measures are monotonically increasing functions of $z$. A more careful examination shows that although $S_E<C_E$ for small values of $z$, in the large squeezing limit the entropy becomes much larger than the capacity. We have plotted an explicit example of this behavior in the left panel of figure \ref{fig:seces2zhalf}. Based on this figure, we see that the entanglement and Renyi entropies go on to grow indefinitely as we increase $z$, while the capacity  of entanglement saturates to a finite value. Remarkably, in \cite{Katsinis:2023hqn} a large $z$ expansion was performed for an arbitrary harmonic chain which shows that at leading order the entanglement entropy is time-independent and proportional to the volume of the smaller subsystem.
Moreover, in \cite{MohammadiMozaffar:2024uyp} a similar approach employed which shows that the capacity of entanglement is also obeys a volume law scaling in large squeezing limit such that $C_E/S_E\sim z^{-1}\ll 1$. The right panel of figure \ref{fig:seces2zhalf} illustrates the mean values of the measures as a function of $z$ for $\widetilde{N}=N/2$ which shows qualitatively similar behavior. 

\begin{figure}[h!]
	\begin{center}
\includegraphics[scale=0.59]{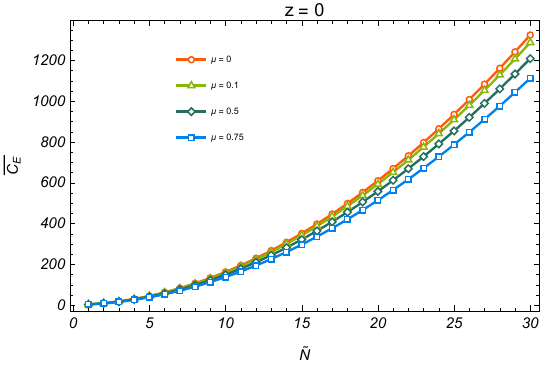}
\includegraphics[scale=0.59]{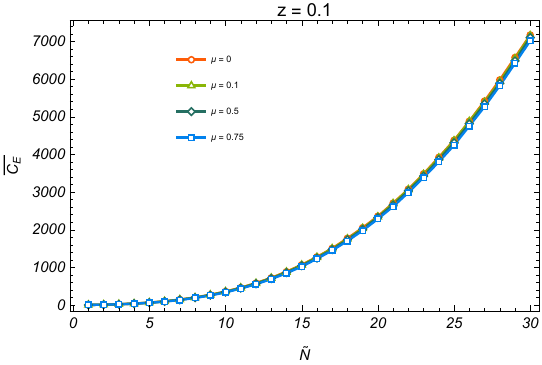}
\includegraphics[scale=0.59]{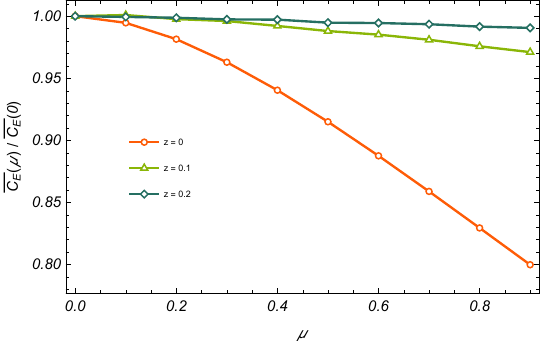}
  	\end{center}
	\caption{Mean value of the capacity of entanglement as a function of $\widetilde{N}$ for several values of $\mu$ and $z$ (left and middle panels). Mean value of the capacity of entanglement as a function of mass for $\widetilde{N}=N/2$ and several values of $z$ (right panel).}
	\label{fig:cmass}
\end{figure}

To close this section, we examine the mass dependence of the capacity of entanglement in figure \ref{fig:cmass}.
The left and middle panels in this figure show the mean value of the capacity of entanglement as a function of $\widetilde{N}$ for several values of $\mu$ and $z$. We see that for massive theory $\overline{C_E}$ decreases as one increases the mass parameter. However, this behavior becomes less pronounced as we increase the squeezing. Hence, the behavior of the massless case at fixed $z$ is the same as the behavior of large $z$ at fixed mass. This is better shown in the right panel that reports the ratio $\frac{\overline{C_E}(\mu)}{\overline{C_E}(0)}$ versus $\mu$ for several values of the squeezing parameter with $\widetilde{N}=\frac{N}{2}$. Clearly, the rate of change of $\overline{C_E}(\mu)$ is a monotonically decreasing function of $z$. Note that in our numerics we have chosen the mass scale to $\mu<1$, which corresponds to a typical correlation length of $\xi\sim \mu^{-1}>1(=\epsilon)$. 
%

\section{Field space entanglement}\label{sec:fieldspace}
In this section we continue our study by computing the field space capacity of entanglement in interacting scalar theories. As we have mentioned before in this case the reduced density matrix can be obtained by summing over a subset of the quantum fields and the resultant entanglement entropy is proportional to the spatial volume. In what follows, we first study the case of two interacting scalar fields and then we consider more general cases with $N$ number of scalar fields with different interaction patterns. 

\subsection{Two interacting scalar fields}
Let us begin by considering the case of two interacting scalar fields whose action is given by eq. \eqref{twofield}. For simplicity we restrict our analysis to Gaussian interactions and thus the corresponding Lagrangian is quadratic. In this case the ground state wave function is Gaussian and most of the expressions can be evaluated analytically which allows us to derive in detail several features of the entanglement measures. Indeed, similar setup first introduced in \cite{Callan:1994py} for computing geometric entropy which can be readily extend to field space as was done in \cite{Mollabashi:2014qfa}. Here we would like to apply the techniques developed in these
references to examine the behavior of field space capacity of entanglement. To do so, consider a general two point function for $\phi$ and $\psi$ as follows (up to a certain normalization factor)
\begin{eqnarray}\label{Psi}
\langle \{\phi,\psi\}|\Psi\rangle\propto \exp\left\{-\frac{1}{2}\int d^{d-1}xd^{d-1}y\left(\phi(x)G_1(x,y)\phi(y)+\psi(x)G_2(x,y)\psi(y)+2\phi(x)G_3(x,y)\psi(y)\right)\right\},
\end{eqnarray}
where $G_{1,2,3}(x,y)$ are complex symmetric functions. The corresponding reduced density matrix for $\psi$ can be evaluated as 
\begin{eqnarray}
\rho_\psi(\psi_1, \psi_2)&=&\int\mathcal{D}\phi \langle \{\phi,\psi_1\}|\Psi\rangle\langle \Psi|\{\phi,\psi_2\}\rangle\nonumber\\ &=&\mathcal{N}\exp\left\{-\frac{1}{2}\int d^{d-1}xd^{d-1}y
\left(\begin{matrix}
\psi_1(x) & \psi_2(x)
\end{matrix}\right)
\left(\begin{matrix}
X(x, y) & 2Y(x, y)  \\
2Y(x, y) & X^*(x, y)
\end{matrix}\right)
\left(\begin{matrix}\psi_1(y) \\ \psi_2(y)
\end{matrix}\right)\right\},\nonumber
\end{eqnarray}
where $\mathcal{N}$ is a normalization constant which can be find by setting ${\rm Tr}\left(\rho_\psi\right)=1$ and
\begin{eqnarray}
X=G_2-G_3(G_1+G_1^*)^{-1}G_3,\hspace*{1cm}Y=-\frac{1}{2}{\rm Re}\left(G_3(G_1+G_1^*)^{-1}G^*_3\right).
\end{eqnarray}
Now using the replica trick and after a mostly straightforward calculation we have
\begin{eqnarray}
{\rm Tr}\rho_\psi^n={\mathcal{N}}^n \exp\left\{-\int d^{d-1}xd^{d-1}y
\left(\begin{matrix}
\psi_1(x),&\cdots, & \psi_n(x)
\end{matrix}\right)
M_n(x, y)
\left(\begin{matrix}\psi_1(y) \\\vdots \\ \psi_n(y)
\end{matrix}\right)\right\},
\end{eqnarray}
where
\begin{eqnarray}
M_n(x, y)=\left(\begin{matrix}
Re X & Y  &0&0&\cdots &Y\\
Y & ReX &Y&0&\cdots &0\\
0 & Y &ReX&Y&\cdots &0\\
\vdots & \vdots &\vdots&\ddots&\vdots &\vdots\\
0 & 0 &0&Y&Re X &Y\\
Y & 0 &0&0&Y &Re X
\end{matrix}\right).
\end{eqnarray}
Further, summing over $\psi$ gives
\begin{eqnarray}
{\rm Tr}\rho_\psi^n=\left(\det \left(\pi^{-1}\left(ReX+2Y\right)\right)\right)^{\frac{n}{2}}\left(\det \left(\pi^{-1}M_n\right)\right)^{-\frac{1}{2}}.
\end{eqnarray}
Inserting the above expression into eq. \eqref{renyi}, the corresponding Renyi entropy becomes \cite{Mollabashi:2014qfa}
\begin{eqnarray}\label{2fieldSn}
S_{n}=\sum_i\frac{n\log(1-\xi_i)-\log(1-\xi_i^n)}{1-n}, \;\;\;\;\;\;\;\;\;\;\;\xi_i=\frac{1-\sqrt{1-z_i^2}}{z_i},
\end{eqnarray}
where
$z_i$ denotes the eigenvalues of $Z\equiv-2Y({\rm Re}\,X)^{-1}$ matrix. Further, combining the above result with eq. \eqref{capa}, we obtain the capacity of entanglement as follows
\begin{eqnarray}\label{2fieldCE}
C_E=\sum_i\xi_i\left(\frac{\log\xi_i}{1-\xi_i}\right)^2.
\end{eqnarray}
So far we have mostly been keeping the discussion at a general level, without specifying the explicit form of the interaction term $\mathcal{L}_{\rm int.}(\phi, \psi)$. In the subsequent sections, we will narrow our focus to specific cases including kinetic mixing and massive interactions. 

\subsubsection{Kinetic mixing interactions}\label{kmi}

The first model we consider is that of two massless scalar fields interacting via a kinetic mixing term
\begin{eqnarray}\label{masslessmodel}
S=\frac{1}{2}\int d^dx\left(\partial_\mu \phi\partial^\mu \phi+\partial_\mu \psi\partial^\mu \psi+\lambda \partial_\mu \phi\partial^\mu \psi\right),
\end{eqnarray}
where $\lambda$ determines the strength of the coupling between the two fields such that the weak and
strong coupling regimes correspond to $\lambda\rightarrow 0$ and $\lambda\rightarrow 2$ limits respectively. Let us recall that the positivity of the corresponding Hamiltonian implies that $|\lambda|\leq 2$. It is relatively simple to show that the ground state wave functional becomes \cite{Mollabashi:2014qfa}
\begin{eqnarray}
\langle \{\phi,\psi\}|\Omega\rangle=\mathcal{N}exp\left\{-\frac{1}{2}\int d^{d-1}x\,d^{d-1}y \,W(x, y)\left(\phi(x)\phi(y)+\psi(x)\psi(y)+\lambda\phi(x)\psi(y)\right)\right\},
\end{eqnarray}
where
\begin{eqnarray}\label{W}
W(x, y)=V^{-1}\sum_{k\neq 0} |k| e^{ik(x-y)},
\end{eqnarray}
and $V$ is the spatial volume. We see that the above expression and eq. \eqref{Psi} will be in complete agreement if we choose 
\begin{eqnarray}
G_1=G_2=\frac{2}{\lambda}G_3=W,\hspace*{1cm}X=(1-\lambda^2/8)W,\hspace*{1cm}Y=-(\lambda^2/16) W.
\end{eqnarray}
Moreover, the eigenvalues of $Z$ can then be evaluated as $z=\frac{\lambda^2}{8-\lambda^2}$
and thus $\xi=\frac{1-\sqrt{1-z^2}}{z}$. Note that based on these results the weak and strong coupling regimes
correspond to $\xi\rightarrow 0$ and $\xi\rightarrow 1$ limits respectively. Now using eqs. \eqref{2fieldSn} and \eqref{2fieldCE} we have
\begin{eqnarray}\label{snsm0}
&&S_n(\lambda)=s_n(\lambda)\sum_{k\neq 0}1, \hspace{2cm}s_n(\lambda)=\frac{n\log(1-\xi)-\log(1-\xi^n)}{1-n},\\
&&C_E(\lambda)=c_E(\lambda)\sum_{k\neq 0}1,\hspace{1.8cm} c_E(\lambda)=\frac{\xi\log^2\xi}{(\xi-1)^2}.
\end{eqnarray}
Clearly, the above measures are divergent which can be regularized by introducing a momentum cutoff. Indeed, as shown in \cite{Mollabashi:2014qfa} considering a $(1+1)$-dimensional compact space with length $L$ and imposing periodic boundary condition we have
\begin{eqnarray}\label{infinitesum}
\sum_{k\neq 0}1\sim \sum_{k\neq 0}e^{-\epsilon |k|} =\sum_{n\neq 0}e^{-\epsilon \frac{2\pi}{L}|n|}=\frac{L}{\pi \epsilon}-1+\cdots,
\end{eqnarray}
and thus the entanglement measures obey the volume law.\footnote{Similarly in higher dimensions one finds that the leading divergent term is proportional to $\left(\frac{L}{\epsilon}\right)^{d-1}$.} Although both the entanglement entropy and capacity of entanglement are sensitive to the UV cutoff, but the ratio is finite and scheme independent. Before examining the full $\lambda$-dependence of the measures, we would like to study their asymptotic behaviors in weak and strong coupling regimes. Remarkably, in the weak coupling regime a perturbative expansion yields
\begin{eqnarray}
s_n(\lambda)=\frac{1}{n-1}\left(n\frac{\lambda^2}{16}-\left(\frac{\lambda^2}{16}\right)^n\right)+\cdots,\hspace*{1cm}
c_E(\lambda)=\frac{\lambda^2}{16}\left(\log\frac{\lambda^2}{16}\right)^2+\cdots,
\end{eqnarray}
which shows that in this limit both quantities vanish. This is consistent with the idea that as $\lambda$
decreases, the reduced density matrix becomes more and more separable. On the other hand, in $2-|\lambda|\ll 1$ limit we obtain
\begin{eqnarray}\label{stronglambda}
s_n(\lambda)=-\frac{1}{2}\log(2-|\lambda|)+\frac{\log n}{n-1}-\log 2+\cdots,\hspace*{1cm}
{c}_E(\lambda)=1-\frac{1}{3}(2-|\lambda|)+\cdots,
\end{eqnarray}
and thus
\begin{eqnarray}
\frac{C_E}{S_E}(|\lambda|\rightarrow 2)\sim \frac{-2}{\log(2-|\lambda|)}\ll 1.
\end{eqnarray}
Interestingly, we see that in this limit $C_E\ll S_E$ which is consistent with the idea that the corresponding reduced
density matrix becomes more and more maximally mixed as one increases the coupling.

\begin{figure}[h!]
	\begin{center}
\includegraphics[scale=0.57]{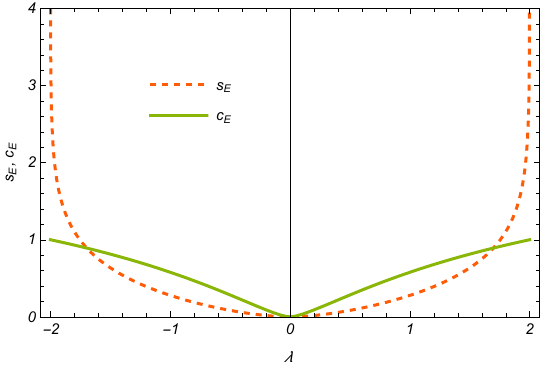}
  \hspace*{0.1cm}
\includegraphics[scale=0.57]{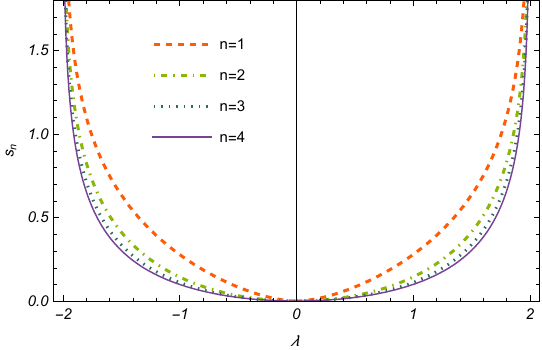}
  \hspace*{0.1cm}
\includegraphics[scale=0.57]{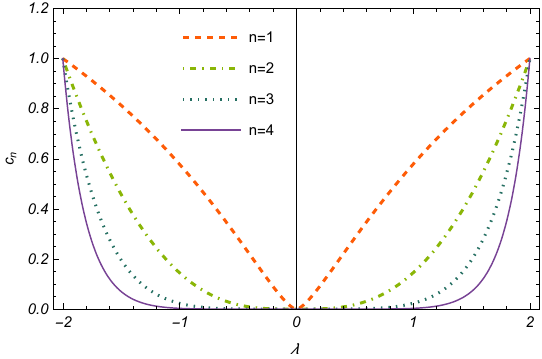}
  	\end{center}
	\caption{$s_E$ and $c_E$ (left), $s_n$ (middle) and $c_n$ (right) as functions of the coupling between the two scalar fields.}
	\label{fig:twoscalarFS1}
\end{figure}

We summarize the numerical results for different entanglement measures as functions of the coupling between the two scalar fields in figure \ref{fig:twoscalarFS1}. The left panel demonstrates the entanglement entropy and capacity of entanglement. We note a number of key features: First, both these measures increase as one increases the coupling.
Second, although the entanglement entropy diverges in the strong coupling limit, the capacity of
entanglement saturates to unity in agreement with eq. \eqref{stronglambda}. Interestingly, for a specific value of
the coupling these measures coincide, i.e., $C_E(\lambda_c)=S_E(\lambda_c)$ where $|\lambda|_c\sim 1.7$. Moreover, the middle panel presents the Renyi entropy as a function of $\lambda$ for several values of $n$. Although $S_n$
is a decreasing function of the Renyi index, it increases with the coupling as expected. In the right
panel we show the $n$-th capacity of entanglement for the same values of the parameters. We see that $C_n$ decreases with $n$ and also saturates from below to unity. From the behavior of $S_n$ and $C_n$, we see that the qualitative dependence of these measures on the coupling is similar to $n=1$ case. Also both measures decrease as we increase the Renyi index. Further, the entanglement entropy and capacity of entanglement have no reason to be equal, except for a certain coupling between the fields. This observation is different from what happens for the geometric entanglement entropy in QFTs with a gravity dual. Indeed, as discussed in \cite{DeBoer:2018kvc} in holographic duals of Einstein gravity, the ratio $C_E/S_E$ turns out to be exactly equal to one.

\subsubsection{Massive interactions}
In this section we evaluate capacity of entanglement and some other entanglement measures for a family of massive interacting scalar theories with the following action \cite{Mollabashi:2014qfa}
\begin{eqnarray}\label{massivemodel}
S=\frac{1}{2}\int d^dx\left(\partial_\mu \phi\partial^\mu \phi+\partial_\mu \psi\partial^\mu \psi-(\phi, \psi)\left(\begin{matrix}
A & C  \\
C & B
\end{matrix}\right)\left(\begin{matrix}
\phi  \\
\psi
\end{matrix}\right)\right),
\end{eqnarray}
where $A, B$ and $C$ are real constants. Using an orthogonal transformation the above action can be diagonalized as follows 
\begin{eqnarray}
S=\frac{1}{2}\int d^dx\left(\partial_\mu \phi'\partial^\mu \phi'+\partial_\mu \psi'\partial^\mu \psi'-m_1^2{\phi'}^2-m_2^2{\psi'}^2\right),
\end{eqnarray}
where
\begin{eqnarray}
\left(\begin{matrix}
\phi'  \\
\psi'
\end{matrix}\right)=
\left(\begin{matrix}
\cos\theta & -\sin\theta  \\
\sin\theta & \cos\theta
\end{matrix}\right)\left(\begin{matrix}
\phi  \\
\psi
\end{matrix}\right).
\end{eqnarray}
Considering the vacuum state for the total Hamiltonian, it is relatively simple to show that in this
case the ground state wave functional is given by eq. \eqref{Psi}, but now $G_i$'s are given by
\begin{eqnarray}
G_i(x, y)=V^{-1}\sum_k \mathcal{G}_{i}(k) e^{ik(x-y)},
\end{eqnarray}
where
\begin{eqnarray}
\mathcal{G}_{1}(k)&=&\sqrt{k^2+m_1^2}\cos^2\theta+\sqrt{k^2+m_2^2}\sin^2\theta,\nonumber\\
\mathcal{G}_{2}(k)&=&\sqrt{k^2+m_1^2}\sin^2\theta+\sqrt{k^2+m_2^2}\cos^2\theta,\nonumber\\
\mathcal{G}_{3}(k)&=&\left(\sqrt{k^2+m_1^2}-\sqrt{k^2+m_2^2}\right)\sin\theta\cos\theta.
\end{eqnarray}
Note that for $m_2=m_1$ we have $\mathcal{G}_{2}=\mathcal{G}_{1}$ and $\mathcal{G}_{3}$ vanishes. Further, the Renyi entropy and capacity of entanglement are obtained by the replacement $z(k)=\left(2\mathcal{G}_1\mathcal{G}_2\mathcal{G}_3^{-2}-1\right)^{-1}$ in eqs. \eqref{2fieldSn} and \eqref{2fieldCE}. Now taking the continuum limit, \textit{i.e.}, $V\rightarrow\infty$
and
$V^{-1}\sum_k(\cdots)\rightarrow(2\pi)^{1-d}\int (\cdots)d^{d-1}k$, one finds
\begin{eqnarray}
S_n&=&\frac{V\Omega_{d-1}}{(2\pi)^{d-1}}\int_0^{\epsilon^{-1}} dk\;k^{d-2}\frac{\log\left(1-\xi^n(k)\right)-n\log\left(1-\xi(k)\right)}{n-1},\nonumber\\
C_E&=&\frac{V\Omega_{d-1}}{(2\pi)^{d-1}}\int_0^{\epsilon^{-1}} dk\;k^{d-2}\xi(k)\left(\frac{\log\xi(k)}{1-\xi(k)}\right)^2,
\end{eqnarray}
where $\epsilon$ is introduced to avoid the ultraviolet divergences. The leading behavior can be found by expanding the above expressions in $k\rightarrow \infty$ limit. Indeed, in this case we have $z\sim\frac{(m_1^2-m_2^2)^2}{32k^4}\sin^2(2\theta)$ which yields
\begin{eqnarray}
S_n&=&\frac{V\Omega_{d-1}}{(2\pi)^{d-1}}\frac{n(m_1^2-m_2^2)^2\sin^22\theta}{64(1-n)}\Bigg\{ \begin{array}{rcl}
&\log \epsilon&\,\,\,d=5,\\
&\frac{1}{(5-d)\epsilon^{d-5}}&\,\,\,d\geq6\\
\end{array}.\nonumber\\
C_E&=&\frac{V\Omega_{d-1}}{(2\pi)^{d-1}}\frac{(m_1^2-m_2^2)^2\sin^22\theta}{4}\Bigg\{ \begin{array}{rcl}
&\frac{-1}{3}\left(\log \epsilon\right)^3&\,\,\,d=5,\\
&\frac{1}{(d-5)\epsilon^{d-5}}\left(\log \epsilon\right)^2&\,\,\,d\geq6\\
\end{array}.
\end{eqnarray}
Also the entanglement entropy is easily found to be \cite{Mollabashi:2014qfa}
\begin{eqnarray}
S_E=\frac{V\Omega_{d-1}}{(2\pi)^{d-1}}\frac{(m_1^2-m_2^2)^2\sin^22\theta}{16}\Bigg\{ \begin{array}{rcl}
&\frac{1}{2}\left(\log \epsilon\right)^2&\,\,\,d=5,\\
&\frac{1}{(5-d)\epsilon^{d-5}}\log\epsilon&\,\,\,d\geq6\\
\end{array}.
\end{eqnarray}
Remarkably, based on the above results we see that the ratio $\frac{C_E}{S_E}$ is scheme independent. Moreover, in all dimensions the volume law scaling is clearly manifest and for $m_2=m_1$, where the fields do not mix with each other, the measures vanish.

\subsection{$N$ interacting scalar fields}
In this section we generalize our studies to $N$ number of massless interacting scalar field theories. Following \cite{MohammadiMozaffar:2015clv}, our calculations here will focus on two specific models with kinetic mixing interactions whose couplings are marginal. The first model is infinite-range model with the following action
\begin{eqnarray}
S_{I}=\frac{1}{2}\int d^dx\left(\sum_{i=1}^{N}\left(\partial_\mu\phi_i\right)^2+\lambda\sum_{1\leq i<j\leq N}\partial_\mu\phi_i\partial^\mu\phi_j\right),
\end{eqnarray}
where we have chosen the same value of coupling between different types of scalar fields for simplicity. The second model is nearest-neighbor model where any field interacts only with its nearest neighbors and the corresponding action is 
\begin{eqnarray}
S_{II}=\frac{1}{2}\int d^dx\left(\sum_{i=1}^{N}\left(\partial_\mu\phi_i\right)^2+\lambda\sum_{\langle i,j \rangle}\partial_\mu\phi_i\partial^\mu\phi_j\right).
\end{eqnarray}
We would like to evaluate the reduced density matrix by tracing over the first $\widetilde{N}$ fields. To do
so, we decompose the total Hilbert space as $\mathcal{H}_{\rm tot.}=\mathcal{H}_{\{\widetilde{N}\}}\otimes \mathcal{H}_{\{N-\widetilde{N}\}}$ where $\mathcal{H}_{\{\widetilde{N}\}}=\mathcal{H}_1\otimes \cdots \otimes \mathcal{H}_{\widetilde{N}}$ and $\mathcal{H}_{\{N-\widetilde{N}\}}=\mathcal{H}_{\widetilde{N}+1}\otimes \cdots \otimes \mathcal{H}_{N}$. In this case the corresponding reduced density matrix is 
\begin{eqnarray}
\rho_{{\{\widetilde{N}\}}}={\rm Tr}_{\{N-\widetilde{N}\}}\rho_{\rm tot.},
\end{eqnarray}
and the Renyi entropy is given by
\begin{eqnarray}
S_n(\widetilde{N}, N)=\frac{1}{1-n}\log {\rm Tr}\rho_{{\{\widetilde{N}\}}}^n,
\end{eqnarray}
which gives the same result as in eq. \eqref{2fieldSn}. Further, the corresponding expressions for entanglement entropy and capacity of entanglement are similar to previous sections. It is relatively straightforward to show that in both infinite-range and nearest-neighbor models the corresponding vacuum state wave functional is given by \cite{MohammadiMozaffar:2015clv}
\begin{eqnarray}\label{wavefunc}
\Psi[\phi_i]=\mathcal{N}\exp\left(-\frac{1}{2}\int d^{d-1}xd^{d-1}y\sum_{i,j=1}^N\phi_i(x)G_{ij}(x, y)\phi_j(y)\right),
\end{eqnarray}
where $G(x, y)=\frac{W(x, y)}{2}\mathcal{G}$ and
\begin{eqnarray}
\mathcal{G}_{IR}=\left(\begin{matrix}
2 & \lambda  &\lambda&\lambda&\cdots &\lambda\\
\lambda & 2 &\lambda&\lambda&\cdots &\lambda\\
\lambda & \lambda &2&\lambda&\cdots &\lambda\\
\vdots & \vdots &\vdots&\ddots&\vdots &\vdots\\
\lambda & \lambda &\lambda&\cdots&2 &\lambda\\
\lambda & \lambda &\lambda&\cdots&\lambda &2
\end{matrix}\right),\hspace{2cm}\mathcal{G}_{NN}=\left(\begin{matrix}
2 & \lambda  &0&0&\cdots &\lambda\\
\lambda & 2 &\lambda&0&\cdots &0\\
0 & \lambda &2&\lambda&\cdots &0\\
\vdots & \vdots &\vdots&\ddots&\vdots &\vdots\\
0 & 0 &0&\lambda&2 &\lambda\\
\lambda & 0 &0&0&\lambda &2
\end{matrix}\right).
\end{eqnarray}
In the above expressions $W(x, y)$ is the same as eq. \eqref{W}. In what follows, we mainly focus on the infinite-range model, because the interesting qualitative features of the measures are independent of the details of the interaction pattern. In this case the positivity of the Hamiltonian implies that $\frac{-2}{N-1}\leq \lambda\leq 2$. Further, in order to find the capacity of entanglement using eq. \eqref{2fieldCE}, we note that the corresponding expression for $\xi$ can be written in a closed form for any $N$ and $\widetilde{N}$ as\footnote{We skip over the details of the calculation and we refer the interested reader to \cite{MohammadiMozaffar:2015clv} for further details.}
\begin{eqnarray}
\xi (\widetilde{N}, N)=\frac{1-\sqrt{1-f(\widetilde{N}, N)^2}}{f(\widetilde{N}, N)},
\end{eqnarray}
where
\begin{eqnarray}
f(\widetilde{N}, N)=\frac{(N-\widetilde{N})Y(\widetilde{N})}{(N-\widetilde{N})Y(\widetilde{N})-(N-\widetilde{N}+1)\lambda-2}\hspace*{1.2cm}Y(\widetilde{N})=\frac{\widetilde{N}}{1+(\widetilde{N}-1)\lambda/2}\left(\frac{\lambda }{2}\right)^2.
\end{eqnarray}
Again the resultant measures are divergent and we can employ a similar method as in eq. \eqref{infinitesum} to regularize the corresponding expressions. In what follows we consider density of these measures, \textit{i.e.}, the measures in units of the infinite volume. It is straightforward to show that in the weak coupling limit a perturbative expansion yields
\begin{eqnarray}
s_E&=&-\frac{\widetilde{N}(N-\widetilde{N})}{16}\lambda^2\log\left[\lambda^2 \widetilde{N}(N-\widetilde{N})\right]+\cdots,\\
c_E&=&\frac{\widetilde{N}(N-\widetilde{N})}{16}\lambda^2\left(\log\left[\lambda^2 \widetilde{N}(N-\widetilde{N})\right]\right)^2+\cdots.
\end{eqnarray}
Further, in $2-|\lambda|\ll 1$ limit we have
\begin{eqnarray}\label{secelargelambda}
s_E=-\frac{1}{2}\log|2-\lambda|+\cdots,\hspace*{1cm}c_E=1-\frac{N}{6\widetilde{N}(N-\widetilde{N})}|2-\lambda|+\cdots,
\end{eqnarray}
which shows that in the strong coupling limit the corresponding reduced density matrix becomes more and more maximally mixed. Note that we consider a pure state and hence all the above expressions are symmetric under $\widetilde{N} \rightarrow N-\widetilde{N}$.

\begin{figure}[h!]
	\begin{center}
\includegraphics[scale=0.78]{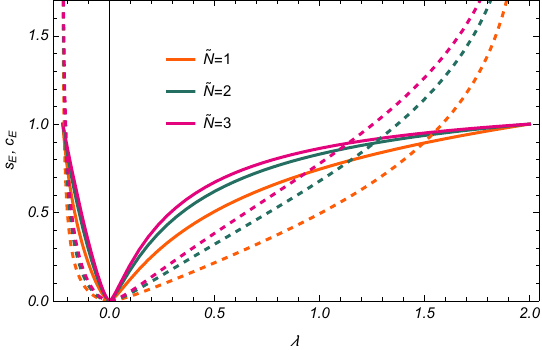}
  \hspace*{0.4cm}
\includegraphics[scale=0.78]{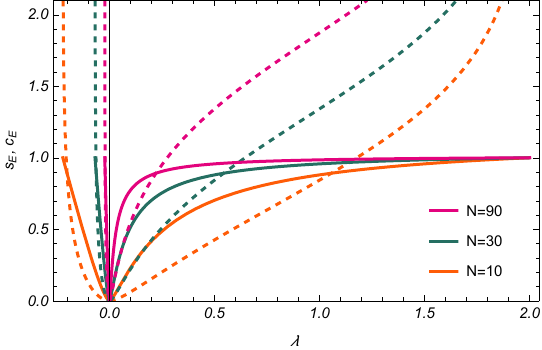}
  	\end{center}
	\caption{Capacity of entanglement (solid) and entanglement entropy (dashed) as functions of the coupling constant for $N=10$ (left) and $\widetilde{N}=\frac{N}{2}$ (right).}
	\label{fig:infiniterange1}
\end{figure}

\begin{figure}[h!]
	\begin{center}
\includegraphics[scale=0.78]{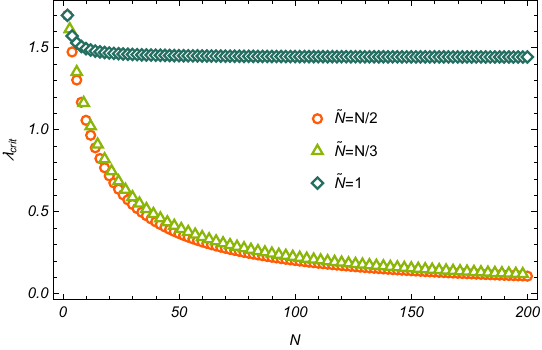}
 \hspace*{0.4cm}
\includegraphics[scale=0.78]{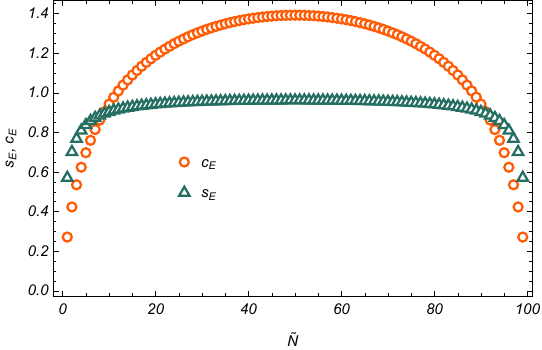}
  	\end{center}
	\caption{\textit{Left}: $\lambda_{\rm crit.}$ as a function of $N$ for several values of $\widetilde{N}$. \textit{Right}: Capacity of entanglement and entanglement entropy as functions of $\widetilde{N}$ for $\lambda=0.5$ and $N=100$. }
	\label{fig:infiniterange2}
\end{figure}

In order to gain further insights into certain properties of these quantities we summarize the full $\lambda$-dependence of the entanglement measures in figures \ref{fig:infiniterange1} and \ref{fig:infiniterange2}. Figure \ref{fig:infiniterange1} presents $s_E$ and $c_E$ as functions of the coupling between the scalar fields. Both these measures increase as one increases the coupling such that the entanglement entropy diverges in the strong coupling limit and the capacity of entanglement saturates to unity which is consistent with eq. \eqref{secelargelambda}. Interestingly, for specific values of the coupling these measures coincide, \textit{i.e.}, $c_E(\lambda_{\rm crit.})=s_E(\lambda_{\rm crit.})$ where the value of $\lambda_{\rm crit.}$ depends on $N$ and $\widetilde{N}$. In the left panel of figure \ref{fig:infiniterange2} we plot $\lambda_{\rm crit.}(N)$ for several values of $\widetilde{N}$. Based on this plot, we see that for large values of $N$, and when $\widetilde{N}\sim \mathcal{O}(N)$, the two measures coincide in the weak coupling limit. 
Further, in the right panel we present entanglement entropy and capacity of entanglement as functions of $\widetilde{N}$ for specific values of $\lambda$ and $N$. The $\widetilde{N} \rightarrow N-\widetilde{N}$ symmetry is clearly manifest.

\begin{figure}[h!]
	\begin{center}
\includegraphics[scale=0.78]{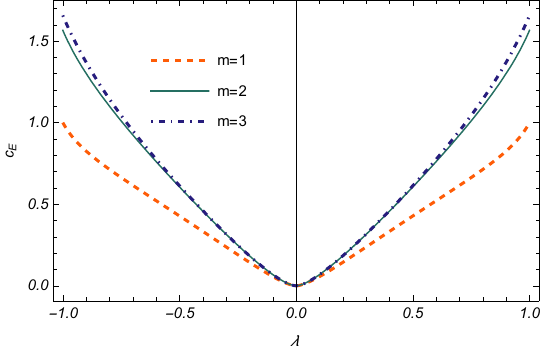}
  \hspace*{0.8cm}
\includegraphics[scale=0.78]{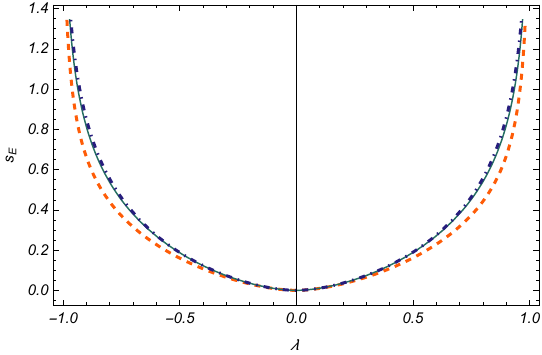}
  	\end{center}
	\caption{Capacity of entanglement (left) and entanglement entropy (right) as functions of the coupling constant for nearest-neighbor model with $N=8$.}
	\label{fig:nearest1}
\end{figure}

At this point, let us turn our attention to the nearest-neighbor model. In this case the positivity of the Hamiltonian implies that $|\lambda|\leq 1$ and $-1\leq \lambda\leq (\cos\frac{\pi}{N})^{-1}$ for even and odd values of $N$ respectively.\footnote{Note that for the special case of $N=2$ this model reduces to the one presented in section \ref{kmi} and we have $|\lambda|\leq 2$.} Again we can find the measures using eqs. \eqref{2fieldSn} and \eqref{2fieldCE}, although in this case one needs a long, involved computation to find the corresponding expression for $\xi$. Here we skip over the details of the calculation and we refer the interested reader to \cite{MohammadiMozaffar:2015clv} for further details. 
It is straightforward to show that in the weak coupling limit we have
\begin{eqnarray}
s_E=-\frac{\lambda^2}{8}\log\lambda^2+\cdots,\hspace{2cm} c_E=\frac{\lambda^2}{8}\left(\log\lambda^2\right)^2+\cdots.
\end{eqnarray}
Figure \ref{fig:nearest1} presents $s_E$ and $c_E$ as functions of the coupling between the scalar fields. Again, both the measures increase as one increases the coupling such that the entanglement entropy diverges in the strong coupling limit and the capacity of entanglement saturates to a finite value. We see that the saturation value of $c_E$ depends on $N$ and $\widetilde{N}$, unlike the infinite-range model. Further, for large values of $N$ and $\widetilde{N}$ these two measures coincide in the weak coupling limit. 

\section{Conclusions and Discussions}\label{sec:diss}

In this paper, we explored several aspects of capacity of entanglement, the quantum information theoretic counterpart of heat capacity, in certain setups whose entanglement entropy obeys a volume law scaling. We have mainly studied the behavior of ``geometric'' capacity of entanglement between two different spatial regions of a field theory or its regularized counterpart on a lattice. We also generalize our study to the case of ``field space'' capacity of entanglement
by considering interacting scalar fields with different interaction patterns. We discuss how our results are consistent with the behavior of other entanglement measures including entanglement and Renyi entropies. In
the following, we would like to summarize our main results and also discuss some further problems.

\begin{itemize}

\item In a two dimensional $p$-alternating sublattice on a periodic lattice, the capacity of entanglement and Renyi entropies obey a volume law and become extensive. In this case discrete translation symmetry along the entangling region was preserved which allows us an analytical treatment. We have studied several aspects of the capacity of entanglement and Renyi entropy for different values of the parameters numerically both in the vacuum and thermal states. Moreover, we have also obtained analytical results in the specific regimes of the parameter space. Specifically, a careful examination shows that in $T\gg m$ limit the corresponding reduced density matrix becomes more and more maximally mixed which implies $C_E\ll S_E$. 

\item In a non-local scalar field theory whose action contains spatial higher derivatives, the ground
state capacity of entanglement follows a volume law as long as the size of the entangling region is smaller than a certain length scale. This scale is related to the number of lattice sites which are correlated together (due to the presence of higher derivative terms) and also on the correlation strength between the lattice points. Moreover, for larger entangling regions, $C_E$ saturates to a finite value. Remarkably, in both regimes the corresponding reduced density matrix becomes more and more maximally mixed as one increases the strength of the correlation between the lattice sites.

\item Considering a $(3+1)$-dimensional free scalar theory in a squeezed state, the corresponding capacity of entanglement monotonically increases with $z$ and obeys a volume law even for small values of the squeezing parameter. A more careful examination shows that although $S_E<C_E$ for small values of $z$, in the large squeezing limit the entropy becomes much larger than the capacity. Moreover, we see that in this limit the capacity of entanglement at leading order is time-independent and saturates to a finite value. We have also observed that for massive theory the capacity of entanglement decreases as one increases the mass parameter. However, this behavior becomes less pronounced
as we increase $z$. Hence, the behavior of the massless case at fixed $z$ is the same as the behavior of large $z$ at fixed mass. 

\item We have also generalized the notion of capacity of entanglement to setups where more than one field lives in a field theory to examine the behavior of field space capacity of entanglement. In particular, considering the case of two interacting massless scalar fields, we have found that $C_E$ increases as one increases the coupling. Moreover, although the entanglement entropy diverges in the strong coupling limit, the capacity of entanglement saturates to unity.
Interestingly, these two measures coincide in the relatively strong coupling limit. On the other hand, in the case of $N$ interacting scalars, we have shown that the interesting qualitative features of $S_E$ and $C_E$ are independent of the details of the interaction pattern. Again, both these measures increase as one increases the coupling such that the entanglement entropy diverges in the strong coupling limit and the capacity of entanglement saturates to a finite value. Further, for a large number of interacting fields these two measures coincide in the weak coupling limit. Indeed, our numerical results make it clear that in the strong coupling limit the corresponding reduced density matrix becomes more and more maximally mixed, \textit{i.e.}, $C_E/S_E\ll 1$. This behavior presumably means maximal entanglement in field space. 

\end{itemize}

Interestingly enough, based on the above results we see that the entanglement entropy and the capacity of entanglement have no reason to be equal in general situations. However, it was argued in \cite{DeBoer:2018kvc} that for quantum field theories dual to classical Einstein gravity $S_E$ and $C_E$ exactly coincide with each other. Thus, at least in these specific models our results suggest that the corresponding states under study may not have a solution of a classical gravity theory as a holographic dual.

There are several interesting directions which one can follow to further investigate the different aspects of capacity of entanglement both in field theory and holography. An interesting question is to explore the existence of consistent holographic duals for states with volume law scaling in the gravity side. Indeed, in our study we have found that $C_E/S_E\rightarrow 0$ in specific regimes of the parameter space which means that the corresponding reduced density matrix can be approximated as proportional to identity. Hence, the corresponding Renyi entropies are independent of the Renyi index and we have a flat entanglement spectra. The situation is similar to holographic fixed-area states which have $n$-independent Renyi entropies \cite{Dong:2018seb,Akers:2018fow}. It is an interesting question to explore possible connections between the states with volume law scaling in the field theory and fixed-area states of quantum gravity.

It is also interesting to study the scaling and time evolution of capacity of entanglement in nonrelativistic theories, in particular those with Lifshitz exponent \cite{Mozaffar:2021nex,prep}. Another way to extend our study is to go beyond the free theories and consider interactions both in relativistic \cite{Hertzberg:2012mn} or nonrelativistic setups \cite{Rabideau:2015via}. Indeed, this analysis may shed more light on the role of interactions on the entanglement structure. Moreover, it will be an important future problem to study the possible connection between the magic and non-flatness of entanglement spectrum which can be characterized by the capacity of entanglement, \textit{e.g.}, see \cite{White:2020zoz,Cao:2024nrx}. \footnote{We would like to thank the anonymous referee for bringing this to our attention.} We leave further investigations of these aspects to future work.

\subsection*{Acknowledgements}
We would like to thank Komeil Babaei Velni and Sare Khoshdooni for ongoing collaboration on related ideas. It is our pleasure to thank Krysztof Andrzejewski and Saleh Rahimi-Keshari for valuable discussions and correspondence. This work is based upon research funded by Iran National Science Foundation (INSF) under project No. 4013637.


\begin{thebibliography}{}

\bibitem{Eisert:2008ur}
J.~Eisert, M.~Cramer and M.~B.~Plenio,
``Area laws for the entanglement entropy - a review,''
Rev. Mod. Phys. \textbf{82}, 277-306 (2010)
doi:10.1103/RevModPhys.82.277
[arXiv:0808.3773 [quant-ph]].



\bibitem{Calabrese:2009qy}
P.~Calabrese and J.~Cardy,
``Entanglement entropy and conformal field theory,''
J. Phys. A \textbf{42}, 504005 (2009)
doi:10.1088/1751-8113/42/50/504005
[arXiv:0905.4013 [cond-mat.stat-mech]].

\bibitem{Laflorencie:2015eck}
N.~Laflorencie,
``Quantum entanglement in condensed matter systems,''
Phys. Rept. \textbf{646}, 1-59 (2016)
doi:10.1016/j.physrep.2016.06.008
[arXiv:1512.03388 [cond-mat.str-el]].


\bibitem{Nishioka:2018khk}
T.~Nishioka,
``Entanglement entropy: holography and renormalization group,''
Rev. Mod. Phys. \textbf{90}, no.3, 035007 (2018)
doi:10.1103/RevModPhys.90.035007
[arXiv:1801.10352 [hep-th]].


\bibitem{Casini:2022rlv}
H.~Casini and M.~Huerta,
``Lectures on entanglement in quantum field theory,''
PoS \textbf{TASI2021}, 002 (2023)
doi:10.22323/1.403.0002
[arXiv:2201.13310 [hep-th]].

%
%
%
%
%






\bibitem{Yao:2010woi}
H.~Yao and X.~L.~Qi,
``Entanglement entropy and entanglement spectrum of the Kitaev model,''
Phys. Rev. Lett. \textbf{105}, no.8, 080501 (2010)
doi:10.1103/PhysRevLett.105.080501
[arXiv:1001.1165 [cond-mat.str-el]].





\bibitem{DeBoer:2018kvc}
J.~De Boer, J.~J\"arvel\"a and E.~Keski-Vakkuri,
``Aspects of capacity of entanglement,''
Phys. Rev. D \textbf{99}, no.6, 066012 (2019)
doi:10.1103/PhysRevD.99.066012
[arXiv:1807.07357 [hep-th]].


\bibitem{Arias:2023kni}
R.~Arias, G.~Di Giulio, E.~Keski-Vakkuri and E.~Tonni,
``Probing RG flows, symmetry resolution and quench dynamics through the capacity of entanglement,''
JHEP \textbf{03}, 175 (2023)
doi:10.1007/JHEP03(2023)175
[arXiv:2301.02117 [cond-mat.stat-mech]].

\bibitem{Nakaguchi:2016zqi}
Y.~Nakaguchi and T.~Nishioka,
``A holographic proof of R\'enyi entropic inequalities,''
JHEP \textbf{12}, 129 (2016)
doi:10.1007/JHEP12(2016)129
[arXiv:1606.08443 [hep-th]].

\bibitem{Nakagawa:2017wis}
Y.~O.~Nakagawa and S.~Furukawa,
``Capacity of entanglement and the distribution of density matrix eigenvalues in gapless systems,''
Phys. Rev. B \textbf{96}, no.20, 205108 (2017)
doi:10.1103/PhysRevB.96.205108
[arXiv:1708.08924 [cond-mat.str-el]].




\bibitem{Okuyama:2021ylc}
K.~Okuyama,
``Capacity of entanglement in random pure state,''
Phys. Lett. B \textbf{820}, 136600 (2021)
doi:10.1016/j.physletb.2021.136600
[arXiv:2103.08909 [hep-th]].


\bibitem{Nandy:2021hmk}
P.~Nandy,
``Capacity of entanglement in local operators,''
JHEP \textbf{07}, 019 (2021)
doi:10.1007/JHEP07(2021)019
[arXiv:2106.00228 [hep-th]].

\bibitem{Bhattacharjee:2021jff}
B.~Bhattacharjee, P.~Nandy and T.~Pathak,
``Eigenstate capacity and Page curve in fermionic Gaussian states,''
Phys. Rev. B \textbf{104}, no.21, 214306 (2021)
doi:10.1103/PhysRevB.104.214306
[arXiv:2109.00557 [quant-ph]].

\bibitem{Wei:2022bed}
L.~Wei,
``Average capacity of quantum entanglement,''
J. Phys. A \textbf{56}, no.1, 015302 (2023)
doi:10.1088/1751-8121/acb114
[arXiv:2205.06343 [math-ph]].

\bibitem{Shrimali:2022bvt}
D.~Shrimali, S.~Bhowmick, V.~Pandey and A.~K.~Pati,
``Capacity of entanglement for a nonlocal Hamiltonian,''
Phys. Rev. A \textbf{106}, no.4, 042419 (2022)
doi:10.1103/PhysRevA.106.042419
[arXiv:2207.11459 [quant-ph]].



\bibitem{Andrzejewski:2023dja}
K.~Andrzejewski,
``Evolution of capacity of entanglement and modular entropy in harmonic chains and scalar fields,''
Phys. Rev. D \textbf{108}, no.12, 125013 (2023)
doi:10.1103/PhysRevD.108.125013
[arXiv:2309.03013 [hep-th]].




\bibitem{Shrimali:2024nbc}
D.~Shrimali, B.~Panda and A.~Pati,
``Stronger Speed Limit for Observables: Tight bound for Capacity of Entanglement, Modular Hamiltonian and Charging of Quantum Battery,''
[arXiv:2404.03247 [quant-ph]].


\bibitem{Ren:2024qmx}
J.~Ren and D.~Q.~Sun,
``Holographic supersymmetric Renyi entropies from hyperbolic black holes with scalar hair,''
[arXiv:2404.05638 [hep-th]].


\bibitem{Banks:2024cqo}
T.~Banks and P.~Draper,
``Generalized Entanglement Capacity of de Sitter Space,''
[arXiv:2404.13684 [hep-th]].


\bibitem{MohammadiMozaffar:2024uyp}
M.~R.~Mohammadi Mozaffar,
``Capacity of entanglement for scalar fields in squeezed states,''
Phys. Rev. D \textbf{110}, no.4, 046021 (2024)
doi:10.1103/PhysRevD.110.046021
[arXiv:2405.09128 [hep-th]].


\bibitem{Shiba:2013jja}
N.~Shiba and T.~Takayanagi,
``Volume Law for the Entanglement Entropy in Non-local QFTs,''
JHEP \textbf{02}, 033 (2014)
doi:10.1007/JHEP02(2014)033
[arXiv:1311.1643 [hep-th]].



\bibitem{Vitagliano:2010db}
G.~Vitagliano, A.~Riera and J.~I.~Latorre,
``Violation of area-law scaling for the entanglement entropy in spin 1/2 chains,''
New J. Phys. \textbf{12}, 113049 (2010)
doi:10.1088/1367-2630/12/11/113049
[arXiv:1003.1292 [quant-ph]].


\bibitem{MohammadiMozaffar:2017nri}
M.~R.~Mohammadi Mozaffar and A.~Mollabashi,
``Entanglement in Lifshitz-type Quantum Field Theories,''
JHEP \textbf{07}, 120 (2017)
doi:10.1007/JHEP07(2017)120
[arXiv:1705.00483 [hep-th]].

\bibitem{He:2017wla}
T.~He, J.~M.~Magan and S.~Vandoren,
``Entanglement Entropy in Lifshitz Theories,''
SciPost Phys. \textbf{3}, no.5, 034 (2017)
doi:10.21468/SciPostPhys.3.5.034
[arXiv:1705.01147 [hep-th]].


\bibitem{Taylor:2015kda}
M.~Taylor,
``Generalized entanglement entropy,''
JHEP \textbf{07}, 040 (2016)
doi:10.1007/JHEP07(2016)040
[arXiv:1507.06410 [hep-th]].


\bibitem{Mollabashi:2014qfa}
A.~Mollabashi, N.~Shiba and T.~Takayanagi,
``Entanglement between Two Interacting CFTs and Generalized Holographic Entanglement Entropy,''
JHEP \textbf{04}, 185 (2014)
doi:10.1007/JHEP04(2014)185
[arXiv:1403.1393 [hep-th]].



\bibitem{MohammadiMozaffar:2015clv}
M.~R.~Mohammadi Mozaffar and A.~Mollabashi,
``On the Entanglement Between Interacting Scalar Field Theories,''
JHEP \textbf{03}, 015 (2016)
doi:10.1007/JHEP03(2016)015
[arXiv:1509.03829 [hep-th]].



\bibitem{Karch:2014pma}
A.~Karch and C.~F.~Uhlemann,
``Holographic entanglement entropy and the internal space,''
Phys. Rev. D \textbf{91}, no.8, 086005 (2015)
doi:10.1103/PhysRevD.91.086005
[arXiv:1501.00003 [hep-th]].




\bibitem{Huffel:2017ewr}
H.~Huffel and G.~Kelnhofer,
``Field Space Entanglement Entropy, Zero Modes and Lifshitz Models,''
Phys. Lett. B \textbf{775}, 229-232 (2017)
doi:10.1016/j.physletb.2017.10.051
[arXiv:1707.00888 [hep-th]].


\bibitem{Nakai:2017qos}
Y.~Nakai, N.~Shiba and M.~Yamada,
``Entanglement Entropy and Decoupling in the Universe,''
Phys. Rev. D \textbf{96}, no.12, 123518 (2017)
doi:10.1103/PhysRevD.96.123518
[arXiv:1709.02390 [hep-th]].



\bibitem{He:2016ohr}
T.~He, J.~M.~Magan and S.~Vandoren,
``Entanglement Entropy of Periodic Sublattices,''
Phys. Rev. B \textbf{95}, no.3, 035130 (2017)
doi:10.1103/PhysRevB.95.035130
[arXiv:1607.07462 [quant-ph]].


\bibitem{Peschel:2002yqj}
I.~Peschel,
``Calculation of reduced density matrices from correlation functions,''
J. Phys. A \textbf{36}, no.14, L205 (2003)
doi:10.1088/0305-4470/36/14/101
[arXiv:cond-mat/0212631 [cond-mat]].


\bibitem{Casini:2009sr}
H.~Casini and M.~Huerta,
``Entanglement entropy in free quantum field theory,''
J. Phys. A \textbf{42}, 504007 (2009)
doi:10.1088/1751-8113/42/50/504007
[arXiv:0905.2562 [hep-th]].



\bibitem{Eisler:2009vye}
V.~Eisler and I.~Peschel,
``Reduced density matrices and entanglement entropy in free lattice models,''
J. Phys. A \textbf{42}, no.50, 504003 (2009)
doi:10.1088/1751-8113/42/50/504003
[arXiv:0906.1663 [cond-mat.stat-mech]].


\bibitem{Vasli:2023syq}
M.~J.~Vasli, K.~Babaei Velni, M.~R.~Mohammadi Mozaffar, A.~Mollabashi and M.~Alishahiha,
``Krylov complexity in Lifshitz-type scalar field theories,''
Eur. Phys. J. C \textbf{84}, no.3, 235 (2024)
doi:10.1140/epjc/s10052-024-12609-9
[arXiv:2307.08307 [hep-th]].

\bibitem{Pirmoradian:2023uvt}
R.~Pirmoradian and M.~R.~Tanhayi,
``Symmetry-Resolved Entanglement Entropy for Local and Non-local QFTs,''
[arXiv:2311.00494 [hep-th]].









\bibitem{Katsinis:2023hqn}
D.~Katsinis, G.~Pastras and N.~Tetradis,
``Entanglement of harmonic systems in squeezed states,''
JHEP \textbf{10}, 039 (2023)
doi:10.1007/JHEP10(2023)039
[arXiv:2304.04241 [hep-th]].


\bibitem{Katsinis:2024sek}
D.~Katsinis, G.~Pastras and N.~Tetradis,
``Entanglement Entropy of a Scalar Field in a Squeezed State,''
[arXiv:2403.03136 [hep-th]].



\bibitem{Bombelli:1986rw}
L.~Bombelli, R.~K.~Koul, J.~Lee and R.~D.~Sorkin,
``A Quantum Source of Entropy for Black Holes,''
Phys. Rev. D \textbf{34}, 373-383 (1986)
doi:10.1103/PhysRevD.34.373



\bibitem{Srednicki:1993im}
M.~Srednicki,
``Entropy and area,''
Phys. Rev. Lett. \textbf{71}, 666-669 (1993)
doi:10.1103/PhysRevLett.71.666
[arXiv:hep-th/9303048 [hep-th]].


\bibitem{Lohmayer:2009sq}
R.~Lohmayer, H.~Neuberger, A.~Schwimmer and S.~Theisen,
``Numerical determination of entanglement entropy for a sphere,''
Phys. Lett. B \textbf{685}, 222-227 (2010)
doi:10.1016/j.physletb.2010.01.053
[arXiv:0911.4283 [hep-lat]].

\bibitem{Kim:2014nza}
N.~Kim,
``On numerical calculation of R\'enyi entropy for a sphere,''
Phys. Lett. B \textbf{733}, 233-236 (2014)
doi:10.1016/j.physletb.2014.04.052


\bibitem{Callan:1994py}
C.~G.~Callan, Jr. and F.~Wilczek,
``On geometric entropy,''
Phys. Lett. B \textbf{333}, 55-61 (1994)
doi:10.1016/0370-2693(94)91007-3
[arXiv:hep-th/9401072 [hep-th]].

\bibitem{Page:1993df}
D.~N.~Page,
``Average entropy of a subsystem,''
Phys. Rev. Lett. \textbf{71}, 1291-1294 (1993)
doi:10.1103/PhysRevLett.71.1291
[arXiv:gr-qc/9305007 [gr-qc]].




\bibitem{Page:1993wv}
D.~N.~Page,
``Information in black hole radiation,''
Phys. Rev. Lett. \textbf{71}, 3743-3746 (1993)
doi:10.1103/PhysRevLett.71.3743
[arXiv:hep-th/9306083 [hep-th]].

\bibitem{Kawabata:2021hac}
K.~Kawabata, T.~Nishioka, Y.~Okuyama and K.~Watanabe,
``Probing Hawking radiation through capacity of entanglement,''
JHEP \textbf{05}, 062 (2021)
doi:10.1007/JHEP05(2021)062
[arXiv:2102.02425 [hep-th]].


\bibitem{Kawabata:2021vyo}
K.~Kawabata, T.~Nishioka, Y.~Okuyama and K.~Watanabe,
``Replica wormholes and capacity of entanglement,''
JHEP \textbf{10}, 227 (2021)
doi:10.1007/JHEP10(2021)227
[arXiv:2105.08396 [hep-th]].


\bibitem{Adesso}
G. Adesso, S. Ragy and A. R. Lee, “Continuous Variable Quantum Information: Gaussian
States and Beyond”, Open Systems and Information Dynamics 21 01n02 1440001 (2014)
[arXiv:1401.4679 [quant-ph]]


\bibitem{Bianchi:2015fra}
E.~Bianchi, L.~Hackl and N.~Yokomizo,
``Entanglement entropy of squeezed vacua on a lattice,''
Phys. Rev. D \textbf{92}, no.8, 085045 (2015)
doi:10.1103/PhysRevD.92.085045
[arXiv:1507.01567 [hep-th]].


\bibitem{Bianchi:2021lnp}
E.~Bianchi, L.~Hackl and M.~Kieburg,
``Page curve for fermionic Gaussian states,''
Phys. Rev. B \textbf{103}, no.24, L241118 (2021)
doi:10.1103/PhysRevB.103.L241118
[arXiv:2103.05416 [quant-ph]].


%
%
%

\bibitem{Hartman:2013mia}
T.~Hartman,
``Entanglement Entropy at Large Central Charge,''
[arXiv:1303.6955 [hep-th]].

\bibitem{Hartman:2014oaa}
T.~Hartman, C.~A.~Keller and B.~Stoica,
``Universal Spectrum of 2d Conformal Field Theory in the Large c Limit,''
JHEP \textbf{09}, 118 (2014)
doi:10.1007/JHEP09(2014)118
[arXiv:1405.5137 [hep-th]].


\bibitem{Caputa:2014vaa}
P.~Caputa, M.~Nozaki and T.~Takayanagi,
``Entanglement of local operators in large-N conformal field theories,''
PTEP \textbf{2014}, 093B06 (2014)
doi:10.1093/ptep/ptu122
[arXiv:1405.5946 [hep-th]].



\bibitem{Asplund:2015eha}
C.~T.~Asplund, A.~Bernamonti, F.~Galli and T.~Hartman,
``Entanglement Scrambling in 2d Conformal Field Theory,''
JHEP \textbf{09}, 110 (2015)
doi:10.1007/JHEP09(2015)110
[arXiv:1506.03772 [hep-th]].













\bibitem{Karczmarek:2013xxa}
J.~L.~Karczmarek and C.~Rabideau,
``Holographic entanglement entropy in nonlocal theories,''
JHEP \textbf{10}, 078 (2013)
doi:10.1007/JHEP10(2013)078
[arXiv:1307.3517 [hep-th]].






\bibitem{Dong:2018seb}
X.~Dong, D.~Harlow and D.~Marolf,
``Flat entanglement spectra in fixed-area states of quantum gravity,''
JHEP \textbf{10}, 240 (2019)
doi:10.1007/JHEP10(2019)240
[arXiv:1811.05382 [hep-th]].

\bibitem{Akers:2018fow}
C.~Akers and P.~Rath,
``Holographic Renyi Entropy from Quantum Error Correction,''
JHEP \textbf{05}, 052 (2019)
doi:10.1007/JHEP05(2019)052
[arXiv:1811.05171 [hep-th]].



\bibitem{Mozaffar:2021nex}
M.~R.~M.~Mozaffar and A.~Mollabashi,
``Time scaling of entanglement in integrable scale-invariant theories,''
Phys. Rev. Res. \textbf{4}, no.2, L022010 (2022)
doi:10.1103/PhysRevResearch.4.L022010
[arXiv:2106.14700 [hep-th]].


\bibitem{prep}
S. Khoshdooni, K. Babaei Velni and M. Reza Mohammadi Mozaffar, in preparation.


\bibitem{Hertzberg:2012mn}
M.~P.~Hertzberg,
``Entanglement Entropy in Scalar Field Theory,''
J. Phys. A \textbf{46}, 015402 (2013)
doi:10.1088/1751-8113/46/1/015402
[arXiv:1209.4646 [hep-th]].


\bibitem{Rabideau:2015via}
C.~Rabideau,
``Perturbative entanglement entropy in nonlocal theories,''
JHEP \textbf{09}, 180 (2015)
doi:10.1007/JHEP09(2015)180
[arXiv:1502.03826 [hep-th]].

\bibitem{White:2020zoz}
C.~D.~White, C.~Cao and B.~Swingle,
``Conformal field theories are magical,''
Phys. Rev. B \textbf{103}, no.7, 075145 (2021)
doi:10.1103/PhysRevB.103.075145
[arXiv:2007.01303 [quant-ph]].


\bibitem{Cao:2024nrx}
C.~Cao, G.~Cheng, A.~Hamma, L.~Leone, W.~Munizzi and S.~F.~E.~Oliviero,
``Gravitational back-reaction is magical,''
[arXiv:2403.07056 [hep-th]].





\end{thebibliography}
\end{document}